\documentclass[preprint]{aastex}
\newcommand{\cldy}{{\it CLOUDY}}

\lefthead{Shore et al.}
\righthead{V382 Velorum 1999}

\begin{document}

\title{The Early Ultraviolet Evolution of the ONeMg Nova V382 Velorum 1999}

\author{Steven N. Shore}
\affil{Dept. of Physics and Astronomy, Indiana University South Bend, 
1700 Mishawaka Ave, South Bend, IN 46634-7111; 
Osservatorio Astrofisico di Arcetri, 5 Largo E. Fermi, 
I-50125 Firenze, Italy; Department of Physics, University of Pisa, 
Via Buonarroti 2, Pisa, 56100 Italy}
\email{sshore@paladin.iusb.edu}

\author{Greg Schwarz}
\affil{Steward Observatory, University of Arizona, Tucson, AZ 85721}
\email{gschwarz@as.arizona.edu}

\author{Howard E. Bond and Ronald A. Downes}
\affil{Space Telescope Science Institute, 3700 San Martin Drive, 
Baltimore, MD 21218}
\email{bond@stsci.edu and downes@stsci.edu}

\author{Sumner Starrfield}
\affil{Dept. of Physics and Astronomy, Arizona State University, Tempe, AZ 
85287-1504}
\email{starrfield@asu.edu}

\author{A. Evans}
\affil{Dept. of Physics and Astronomy, Keele University, UK}
\email{ae@astro.keele.ac.uk}

\author{Robert D. Gehrz}
\affil {Department of Astronomy, 
University of Minnesota, 116 Church Street, SE, Minneapolis, MN 55455} 
\email{gehrz@hal.astro.umn.edu}

\author{Peter H. Hauschildt}
\affil{Hamburger Landsternwarte, Gojenbergweg 112, 21029 Hamburg, Germany}
\email{peter.hauschildt@hs.uni-hamburg.de}

\author{Joachim Krautter}
\affil{Landessternwarte Heidelberg, Germany}
\email{j.krautter@lsw.uni-heidelberg.de}

\author{Charles E. Woodward}
\affil {Department of Astronomy, 
University of Minnesota, 116 Church Street, SE, Minneapolis, MN 55455} 
\email{chelsea@astro.umn.edu}

\begin{abstract}

{\bf DRAFT} 26/11/02\\
\medskip

We present a multiwavelength study of the ONeMg Galactic nova V382
Velorum 1999 using HST/STIS\footnote{Based on observations made with
the NASA/ESA Hubble Space Telescope, obtained at the Space Telescope
Science Institute, which is operated by Associated Universities for
Research in Astronomy, Inc., under NASA contract 5-26555.  These
observations are associated with proposals for proposal GO 8540 and GO
8671.  Also based on observations made with the NASA-CNES-CSA Far
Ultraviolet Spectroscopic Explorer. FUSE is operated for NASA by the
Johns Hopkins University under NASA contract NAS5-32985.} and FUSE
ultraviolet spectra and comparisons with published groundbased optical
spectra.  We find a close match to the basic phenomenology of another
well-studied ONeMg nova, V1974 Cygni  (Nova Cygni 1992), in particular
to the spectral development through the start of the nebular phase.
Following an ``iron curtain'' phase, the nova proceeded through a
stage of P Cygni line profiles  on all important resonance lines, as
in many ONeMg novae and unlike the CO class. Emergent emission lines
displayed considerable structure, as seen in V1974 Cyg, indicating
fragmentation of the ejecta at the earliest stages of the outburst. 
Analysis and modeling of our ultraviolet spectra suggest that 4 - 5
$\times$ 10$^{-4}$M$_{\odot}$ of material was ejected and that the
distance to the nova is $\simeq 2.5$~kpc. Relative to solar values, we
find the following abundances: He = 1.0, C = 0.6$\pm$0.3, N =
17$\pm$4, O = 3.4$\pm$0.3, Ne = 17$\pm$3, Mg = 2.6$\pm$0.1, Al =
21$\pm$2, and Si = 0.5$\pm$0.3. Finally, we briefly draw comparisons
with Nova LMC 2000, another ONeMg nova, for which similar data were
obtained with HST and FUSE.

\end{abstract}


\section{Introduction} 

Nova Velorum 1999 (V382 Vel) was discovered in outburst by P. Williams,
and independently by C. Gilmore, on 1999 May 20.6 UT \citep{Lee99}.  It
reached a maximum visual magnitude of $V = 2.6$ \citep{SCC99} after a
few days of rise, making it among the brightest novae of the 20th
century.  Its decline showed $t_2 = 4$ days and $t_3 = 10$ days
\citep{DV99} marking it as a ``fast'' nova in the nomenclature
introduced by \citet{Pay57} ($t_2$ and $t_3$ are the times after
maximum for declines of two and three magnitudes, respectively).
\citet{P00} have determined a preoutburst mean V magnitude of 16.$^m$6
and a mean B-V of 0.$^m$14; a pre-outburst B magnitude of 16.4 was
reported \citep{SCC99}. Early optical spectroscopic observations
displayed iron emission associated with the optically thick phase of a
relatively massive ejection, although He I emission was reported
relatively early (2 June, Hidayat et al. 1999).  The first optical
spectra, within two days of the first report, showed P Cyg absorption
components on the Balmer lines \citep{Lee99}. The report of very early
[O III] emission was later corrected, identifying the emission as Fe
II.   Infrared observations \citep{Wood99}  detected the [Ne II]
12.8$\mu$  emission line characteristic of the ``neon nova'' group and
subsequently V382 Vel has been recognized as an ONeMg nova. 

V382 Vel displayed comparable optical phenomenology to the well
studied the ONeMg nova V1974 Cygni ({\it e.g.} Shore et al. 1993, 1994;
Shore 2002) and, as we will show here, its ultraviolet spectrum
developed along similar lines.  We are therefore able to make some
interesting comparisons despite being unable to follow V382 Vel with
the dedicated instrumentation used in our earlier investigations.  In
this paper, we will show that much of what we have learned about the
physics of the early nova outburst from the detailed study of a single
ONeMg nova, V1974 Cyg, is robust: the resemblance between these
outbursts is striking and an important observational constraint on any
model for the nova phenomenon. 

\section{Observations} 

Optical observations have been described by \citet{DV02}. The
ultraviolet (UV) observations reported here were obtained under a
Director's Discretionary target of opportunity program (GO 8540) with
the {\it Space Telescope Imaging Spectrograph} (STIS) on board the
{\it Hubble Space Telescope} (HST) in three epochs using the E140M and
E230M gratings.  Additional spectra were obtained with the {\it Far
Ultraviolet Spectrographic Explorer} (FUSE) satellite on a galactic
nova target of opportunity program at three epochs during 2000 using
the large aperture in both the SiC and LiF channels. These
observations were not accompanied by STIS spectra and were well into
the optically thin stage. The last spectrum contains significant
atmospheric emission and has not been used in this study.  The log of
observations is given in Table 1. 

All spectra have been reduced using standard procedures for both STIS
and FUSE and with software we have previously developed to analyze UV
spectra of novae.  For consistency, we have attempted to duplicate
previous analyses.  Since the FUSE spectra were taken with the large
science aperture, some shifting was required for wavelength
assignments: the STIS and FUSE spectra were cross correlated in
wavelength using the N I multiplet at 1199\AA.  The individual STIS
spectra were registered using several interstellar absorption
features, in particular the C II 1334\AA, Al II 1671\AA, and Mg II
2800\AA\ doublets.  The largest wavelength shift seen in the STIS data
is 18 km s$^{-1}$, while the FUSE spectra were displaced by +40 km
s$^{-1}$ from the STIS data.  After correction, the individual line
profiles were compared in velocity using the \citet{Mor91} and NIST Atomic 
Spectra Database (ASD).\footnote{Accessible through URL: 
$\url{http://physics.nist.gov/cgi-bin/AtData/main_asd}$.} 

\section{Temporal Development of V382 Vel within the First 16 Months} 
\subsection{Spectral Development}

The gallery of merged binned (1\AA\ resolution) spectra is shown in
Figs. 1, 2, and 3, and the high resolution data are displayed for
the 1200-1720\AA\ region in Figs. 4, 5, and 6. Both galleries are
uncorrected for extinction. 

The May 31 (O5JV01) observation occurred during the completely opaque
phase of the ``iron curtain''.  From this stage, based on our previous
experience with V1974 Cyg and related ONeMg novae, we attempted to
predict exposure times on the basis of the rate of development.  The
expansion velocity inferred from the width of the 1700\AA\
pseudo-emission feature was approximately 4000 km s$^{-1}$ and, based
on the the optical light curve, this nova appeared to be developing
about 30\% faster.  Subsequent observations confirmed this behavior:
it was always possible to find an identical phase for each spectrum to
that of V1974 Cyg by scaling the time after optical maximum.  
From this point alone it is clear that the ejecta of the two
novae were similar in dynamics and mass distribution that scaled with
the energy of the outburst. 

The second set of observations, June 22 (O5JV02), displayed strong P
Cygni absorption troughs on many of the usually occurring strong
resonance line profiles, as shown in Fig. 7. This stage is
characteristic of ONeMg novae but not for classical novae of the CO
type.  We should add that two recurrent novae, U Sco during the 1979
outburst, and LMC 1990 \# 2, also displayed strong P Cyg profiles on
ultraviolet resonance lines of Si IV and C IV in IUE spectra.  We will
return to this point below in our discussion of Nova LMC 2000. The
strongest absorption was found for Si IV 1400\AA, which displayed a 
terminal velocity of -5200 km s$^{-1}$.  This radial velocity is substantially
larger than any obtained from the optical line profiles reported by
\citet{DV02} for which the velocities are in closer
agreement with those measured on emission lines observed during the 
August 29 observations (see below, sect. 3.2). In particular, H$\alpha$ on 
31 May showed a
weak P Cyg absorption at -2500 km s$^{-1}$, and low intensity red wing
emission extending to -4000 km s$^{-1}$.  This weak absorption had
disappeared by 25 June, as had the extended wings on both sides of the
line, being replaced by a nearly symmetric profile with FWZI of 4000
km s$^{-1}$. The Al II 2669\AA\ and N IV 1486\AA\ lines at this later
epoch, obtained from our STIS spectrum, showed a nearly identical
profile.  There was still, however, significant overlying line
absorption that likely altered some of the emission line
characteristics.  In particular, the He II 1640\AA\ emission is
flanked by numerous low ionization absorption features and also the
wing of the emerging O III] 1667\AA\ line.  The Mg II profile shows a
weak absorption feature at this stage extending about -4000 km
s$^{-1}$ with a deepest absorption at about -2500 km s$^{-1}$.  The Mg
II profile at the ``matching epoch'' of V1974 Cyg (see below), the
spectrum LWP22786 (see Shore et al. 1993), displays a trough with
deepest absorption at about -3000 km s$^{-1}$.  None of the STIS
spectra were taken early enough to reveal the strong P Cyg phase of
this resonance doublet.  

The P Cyg profiles also provide a clue to the origin of the
photometric and spectroscopic scaling between V382 Vel and V1974 Cyg.
The ratio of the maximum expansion velocity, derived from the P Cyg
profiles was about 1.3 at the same epoch.  It appears the luminosities
of the central stars and the ejecta masses are about the same and that
the scaling results from simply from the relative rate of decrease of
the ejecta column density.  The strength of the P Cyg absorption
trough and the saturation of the profile, especially for the Si IV
1400\AA\ lines, argues for a large covering fraction at this stage of
expansion for the optically thick material.  As we will describe 
below, the later (nebular) stages display line profiles are more
consistent with an axisymmetric than spherical geometry for the
ejecta.  Therefore, as we found with V1974 Cyg and other ONeMg novae,
the early optically thick stages reveal a different ejecta geometry
than the slower moving material observed during the nebular stage. 

The last observation with STIS occurred on August 29.  By this time, the
nova was in the nebular stage.  Strong emission dominated the
1200-2000\AA\ region, especially the resonance lines.  Notably, [Ne
IV] 1602\AA\ was approximately half the intensity of He II 1640\AA\ and
displayed a nearly identical profile, but [Ne V] 1575\AA\ was not
observed.  High ionization species included N V 1240\AA, Si IV + O IV
1400\AA\, and C IV 1550\AA, but there is no trace of O V 1375\AA.  The N IV
1486\AA\ line was strong but there is no visible emission at N IV] 1718\AA.
 Resonance intercombination lines were strong, O III] 1667\AA\ and Si
III] 1895\AA\ and C III] 1909\AA\ being examples.  Several
comparatively low ionization species were still present, including O I
1300\AA, C II 1335\AA, N II 2145\AA, and Mg II 2800\AA.  The FWHM for
all these lines was about 4000 km s$^{-1}$, and the profiles were
nearly identical (see discussion below). 

\subsection{Energetics and Reddening} 

Using $t_2$ = 6 days and $t_3$ = 10 days, \citet{DV02}
derive $M_V({\rm max}) = -8.7\pm 0.2 (1\sigma)$ mag, which translates
to $L_{\rm max} \approx 2\times 10^5$L$_\odot$, and a distance of
about 2 kpc. The first STIS observation occurred within one week of
maximum visual brightness, at a time corresponding approximately to
$t_2$.  The second and third STIS observations were obtained long
after visual maximum, by which time the flux maximum had clearly
shifted into the ultraviolet. They permit an independent determination
of the energetics of the outburst. 

The spectrum obtained on May 21 
most closely resembles the IUE spectrum SWP44156 of
V1974 Cyg 1974 at about 20 days after optical maximum, the uncorrected 
ratio being a factor of 5 (Fig. 8) virtually independent of wavelength.  The
close correspondence of the spectra and the nearly independent flux
ratio suggests the reddening for the two novae is similar and we will
subsequently adopt E(B-V)=0.2 for V382 Vel in the analysis to
follow.\footnote{We note that the \citet{Austin96} value of
E(B-V)=0.3 for V1974 Cyg is likely too large.  A re-analysis of the
spectrum is in preparation, but Draine \& Tan (2002) find that
adopting E(B-V)=0.19 for V1974 Cyg produces good agreement for a model
of of the X-ray scattering halo around this nova.  In the present
analysis, the {\it range}  permitted for the reddening is 0.2 to 0.3.} 
 Only Mg II 2800\AA\ was observed in emission during the first STIS
spectrum.  Its velocity width is consistent with that observed at
H$\alpha$.  For a constant (positive) velocity gradient, the Mg II
velocity indicates the line profile was formed from slower moving gas
situated deeper in the ejecta than the region from which the resonance
absorption trough, on the later-observed P Cyg lines, form.  In
support of this interpretation, we note that in the May 21 spectrum,
we detect a weak P Cyg absorption feature at about -3000 km s$^{-1}$,
consistent with the reported blueward Balmer line 
absorption velocities from the
ESO spectra \citet{DV02}. 

For the June 21 observation, the integrated flux was 6.98$\times
10^{-8}$ erg cm$^{-1}$s$^{-1}$ from 1170 - 3070\AA, uncorrected for
extinction.  The comparison is shown in Fig. 9 with V1974 Cyg (SWP
44378, taken about 50 days past optical maximum). Again, the flux ratio
between the spectra is a factor of 5 in the region shortward of
1700\AA, uncorrected for extinction. For August 29, the flux in this
spectral range was 1.05$\times 10^{-8}$ erg cm$^{-1}$s$^{-1}$.  At
this stage, an approximate match is provided to V1974 Cyg with
spectrum SWP 44378 taken 196 days after optical maximum.  There are,
however, significant differences that are clearly not the result of
extinction.  V382 Vel continued to display a strong O I $\lambda$1302
emission even into the nebular phase and C IV $\lambda$1550 was also
stronger relative to the nitrogen lines.  Al III $\lambda$1860
remained stronger, and while the Si III]/C III] ratio is about the
same as V1974 Cyg, the lines display considerably more knotted
structure.  This may be due, in part, to the resolution of the IUE low
dispersion data but that cannot explain all the differences (see
Fig. 9). 

The interstellar Ly$\alpha$ profile provides additional information on
the possible reddening, yielding a neutral hydrogen column density of
$N_H \approx 1.2\times 10^{21}$cm$^{-2}$.  For the August 29 spectrum,
displayed in Fig. 10a, we assumed a Gaussian profile for the ejecta
emission with a FWHM of $\approx$1500 km s$^{-1}$, scaled to the blue
wing of the observed emission.  At this epoch, there should have been 
no P Cyg absorption trough.  The red wing of Ly$\alpha$ 
is blended with the N V 
profile producing the obvious discrepancy but we cannot obtain an 
unblended, unabsorbed N V profile with which to precisely model the 
interstellar absorption.  Therefore, we concentrate on 
the blue wing of Ly$\alpha$.  It is clear the neutral
hydrogen column density is high, this is supported by the strong H$_2$
absorption seen in the FUSE spectra (see discussion, below) and the
strength of the interstellar lines (Table 5).  Scaling 
E(B-V) = N$_H/3.6\times 10^{21}$cm$^{-2}$ (Savage \& Mathis
1979) yields E(B-V) $\approx 0.3$. 

The first observations of V382 Vel (O5JV01) provide superb high
quality interstellar line profiles from a wide range of absorbers.  
These include the CO (2-1) overtone transitions and, in the FUSE 
spectra, many H$_2$ lines.  While a complete
analysis of these data is beyond the scope of the present paper, a 
study is in preparation, we note here a kinematic constraint on the 
distance to this nova.  An important feature of these spectra is that
the Vela region is especially well observed in high resolution with
both H I and $^{12}$CO (Burton 1985; Dame et al. 1987, 1999). The
Galactic rotation curve determined by Brand \& Blitz (1993) and
standard stars observed in this direction yield a mean $v_{\rm LSR}
\approx -20$ km s$^{-1}$.  There is, however, a particularly
interesting feature of the $^{12}$CO maps, a spur at large positive
LSR velocity, +20 km s$^{-1}$, due to the Carina arm that is also
present in the stronger interstellar lines in V382 Vel, requiring a
lower limit on the distance of about 2 kpc.  Sample profiles are
displayed in Fig. 10b. 

Assuming E(B-V)=0.2, the measured continuum fluxes for the three STIS
observations are 3.67$\times 10^{-7}$ (May 31), 3.02$\times 10^{-7}$
(June 21), and 4.56$\times 10^{-8}$ erg s$^{-1}$ cm$^{-2}$ (August 29)
from 1170-3100\AA.  At the time of the first observation, the optical
flux, based on published UBV photometry, was 5$\times 10^{-8}$ erg
s$^{-1}$ cm$^{-2}$.  Even the first STIS observation shows that the UV
corresponded to most of the emitted flux.  This yields a total
luminosity in the observed band of $> 4.9\times 10^4$L$_\odot$ for a
distance of at least 2 kpc.  
If the distance is increased to 2.5 kpc, this luminosity
becomes nearly identical with V1974 Cyg, about 8$\times
10^4$L$_\odot$, about the Eddington luminosity of a 1.4 M$_\odot$ 
white dwarf (WD) 
\citep{Sho94}.  In their study, Della Valle et al. (2002) assumed
virtually no reddening.  For E(B-V)=0, we find integrated fluxes of
9.57$\times 10^{-8}$, 6.92$\times 10^{-8}$, and 1.05$\times 10^{-8}$
erg s$^{-1}$ cm$^{-2}$, in the respective STIS spectra, that are
incompatible with the distance and spectral comparison with V1974 Cyg.
 The flux ratios between the short wavelength spectra of the two novae
(see Fig. 9), assuming at least as great an extinction for V382 Vel as
V1974 Cyg, gives a distance of 2 kpc assuming a 3.1 kpc distance for
V1974 Cyg (see Paresce et al. 1995). 

We can place independent constraints on the reddening using 
the quiescent luminosity.  This comes from the pre-outburst
observations using the parameters we have derived.  Post-outburst GHRS
spectra of V1974 Cyg revealed a white dwarf with $T_{\rm eff} \approx
2\times 10^4$K after 3 years.  While we do not know if an accretion
disk was established by that time, it is likely one was present for
V382 Vel before outburst.  Taking V = 16.6 and assuming a distance of
2 kpc, the visible colors alone give $L \approx 0.3$L$_\odot$.  For an
accretion disk with a $\nu^{1/3}$ spectral energy distribution, this
becomes $L > 2$L$_\odot$ longward of 1200\AA.  The luminosity is not
unusual for novae entering the later stages of outburst, and is higher
than the last GHRS spectrum we obtained for V1974 Cyg. 

We summarize our distance determinations as follows.  From the maximum 
magnitude - rate of decline (MMRD) 
relation, Della Valle et al. (2002) obtain a distance of 2 kpc.  Based
on comparisons with V1974 Cyg and Nova LMC 2000 (see below), we obtain the
range 2 to 3 kpc.  Interstellar lines constrain the nova to be at
least at the distance of the Carina arm, so $> 2$ kpc.  The
preoutburst luminosity yields 2 kpc while $L_{\rm max} \le L_{\rm
Edd}$ gives 2.5 kpc for a Chandrasekhar mass white dwarf. 

\subsection{Line Profiles}

The first emission lines to appear were the strongest permitted and 
intercombination transitions, 
O I 1302\AA, N II] 2145\AA, Al II 2675\AA, and Mg II
2800\AA.  These showed identical profiles to the optical transitions.
As described by \citet{Hay96} and \citet{Sho98}
for V1974 Cyg, the optical lines suffer less absorption within the
ejecta and deeper layers are observed first at longer wavelengths. The
consistency of the structure over time indicates that the observed
emission knots must have formed early in the outburst -- most likely
at the time of ejection. 

Balmer line profiles obtained on 25 June show strongly asymmetric
structure, with a well defined peak at +800 km s$^{-1}$ and an
uncorrected H$\alpha$/H$\beta$ ratio of about 6.8 \cite{DV02}.
Assuming case A recombination for the highest velocity -- and
presumably most transparent -- portions of the line profile, this
corresponds to E(B-V)$\approx$0.2. This is consistent with our
other determinations.  Several distinct knots appear on all three
principal Balmer profiles, at +200, +400 and +800 km s$^{-1}$ with a
weak extended feature at approximately rest (observer's frame).  No
corresponding knots are seen on the blueshifted side of the profile.
Broad low intensity wings appear on all three profiles, the broadest
is H$\delta$ extending to HWZI of 4000 km s$^{-1}$ while the other two
lines show only around 200 km s$^{-1}$ (Fig. 11). 

A curious feature of the Balmer line profile development is the change
in symmetry between the two epochs, separated by only about 30 days.
The spectra in the first observation are almost identical to the low
ionization inter-system lines observed in the June 21 spectrum.  For
instance, Al II $\lambda$2675 with stronger emission on the
blueshifted side of the profile, while the later spectra do not
resemble any of the UV profiles.  Later spectra are nearly symmetric. 
We note that a comparison of these low ionization profiles with the
reported detection of Li I 6708\AA\ \citep{DV02} suggests
that the latter is likely some other low ionization emission centered
at around 6705\AA.  Given the nitrogen enhancement seen in this nova,
a likely candidate is the doublet N I ($^4$P$^o$ - $^4$D) 6704.84,
6706.11\AA. 

The far ultraviolet emission spectrum was sparse.  We show in Fig. 12 the 
region of Ly$\beta$ and the O VI 1031, 1036\AA\ doublet.  These were 
the only strong emission lines detected in the FUSE spectra, taken within 1.5
years of outburst.  The O VI doublet had a profile that was very 
similar to the optically thin C IV 1550\AA\ doublet which, as we found 
for V1974 Cyg, is consistent with the combined optically thin 
profiles of the doublet 
components whose intrinsic form is similar to the singlets seen in the 
STIS spectra (this is also seen in the STIS profile of the O III] 
1667\AA\ multiplet).  The O VI doublet showed strong decrease
between the 2000 February and 2000 July spectra, dropping from  9.3$\times
10^{-12}$ erg s$^{-1}$cm$^{-2}$ to 2.0$\times 10^{-12}$ erg
s$^{-1}$cm$^{-2}$ (A09303/4).  Both these fluxes are uncorrected for
extinction but have been corrected for line absorption.  The FWHM
remained about 1000 km s$^{-1}$.  The observed decline is completely
consistent with the expected $t^{-3}$ power law for the emission (a
factor of about 3.9) from a freely expanding shell.  No Lyman series
emission lines were seen in either FUSE observation; recall that Ly$\alpha$
showed a strong P Cyg profile in 1999 June but this emission was
absent in the 1999 August observation. 

The line profiles are virtually identical for all components of even
the blended multiplets by the third STIS observation, adding weight to
the assertion that the ejecta were optically thin (nebular) by this
stage.  We compare the profiles of He II $\lambda$1640 and [Ne IV]
$\lambda$1602 in Fig. 13.  Notice that the near identity of the
profiles also argues for chemical homogenization of the ejecta during
the explosion.  There are no indications of the deviations we found
for V1974 Cyg among the individual knots and these knots in the UV
line profiles can also be identified between wavelength regions and at
different epochs.  We remark, however, that these are {\it not}
spatially resolved and the integrated large aperture spectra for V1974
Cyg also did not reveal large deviations.  For example, the Balmer emission
lines were more asymmetric in V382 Vel within the first 3 weeks after outburst 
than the UV lines observed later with STIS.  They quickly transformed, 
however, to the same emission structure (by June 2) that was observed
almost 3 months later in the UV and on other optical lines.  The knot
at +850 km s$^{-1}$ is particularly strong in both the Balmer and UV
lines until 25 June. This is not unexpected, since the ejecta expand
hypersonically and individual knots had not yet recombined by the
third STIS observation.  The later STIS profiles are more symmetric than 
those seen in the first observations in the optical and with STIS. 

In an attempt to determine more information about the structure we
performed a Monte Carlo simulation as described in \citet{Sho93}.  
Two geometries were assumed: a spherical shell, and a thin
ring.  For each, a linear velocity law was assumed and the profile was
rebinned in the observer's frame.  The maximum velocity was fixed at
5200 km s$^{-1}$ using the P Cyg profiles on the UV resonance lines
whose terminal velocities {\it always} exceeded those of the optical
Balmer lines and the later emission line profiles. Figure 14 shows the 
comparison of a sample model profile with 
N IV] 1486\AA\ and He II $\lambda$1640.  Both are assumed
to be an optically thin recombination transitions.  The model profile 
can also be compared with other lines in Fig. 6.  
The line is mainly formed from comparatively low
velocity gas so we explored a range of maximum velocities for the
model.  The good agreement was found for a spherical geometry with
$\Delta R/R = 0.7$ and the density varied as $n(R) \sim R^{-3}$ for a
constant shell mass.  For a ring, almost the same profile is obtained
for $\Delta R/R = 0.5$ for an inclination of 25$^\circ$.  There is a
near degeneracy between the  inclination and thickness for a ring, but
the relative weakness of the extended wings on the ring profile
suggest that the spherical case more closely matches the data.
However, we venture the suggestion that V382 Vel may, when spatially
resolved, contain an elliptical ring with a transverse expansion rate
of 0.2 arcsec yr$^{-1}$ for a distance of 2.3 kpc.  Our model is
derived {\it for this optically thin stage}.  There must be
additional, rapidly expanding matter -- as we found for V1974 Cyg --
in a more spherical distribution to account for the broad shallow
wings observed on all emission line profiles in the earlier spectra. 

A range of inclinations can be estimated using the observed outburst
amplitude and the range in the determined distance. The absolute
quiescent magnitude, uncorrected for inclination, for very fast novae
is 3.76.  At distances between 2 and 3 kpc, the inclination range
required to produce the observed, apparent quiescent magnitude lies
between 45 and 67 degrees assuming an E(B-V) = 0.25.  This differs from the
value obtained by line profile analysis so imaging the spatially
resolved ejecta should decide this issue. 

As seen from Fig. 14, the fit seems to be quite good but since the models
``knots'' are randomly generated, this comparison suggests that there
is nothing particularly informative in the distribution of the knots,
that they are not related directly to the large scale structure of the
ejecta. Rather, we are likely seeing the frozen remnants of an
instability that produced them early in the outburst.  A wind-like
velocity law, $v \sim (1-R_\star/r)^\beta$, where $R_\star$ is the
stellar radius and $\beta$ is a constant, fails to reproduce the line
profile, supporting the contention that the ejecta are freely
expanding and not a wind, at least at the later stages.  This does not
rule out possible wind-ejecta interactions as a source for hard X-ray
emission observed early in the outburst (Mukai \& Ishida 2001, Orio et
al. 2001).  The bulk of the line emission comes from the innermost
portion of the ejecta.  This is also true for the optical line
profiles, which in general sample the denser parts of the ejecta at an
earlier time than the UV (see e.g. Hayward et al. 1996). While a
detailed model is beyond the scope of this paper, we note that the
resemblance of the UV lines in the third STIS spectrum with the first
optical data suggests they are formed in the same part of the ejecta
and could be used for detailed modeling.

It is interesting also to note the weak dependence of profile on
ionization state.  The Ne IV] 1602\AA\ line may have a contribution
from a more spherical distribution (see \cite{Sho93} for discussion),
although the peaks in the line core are indicative of a mainly an
axisymmetric geometry, while He II 1640\AA\ and N IV] 1486\AA\ wings
are narrower and likely formed in a predominantly ring-like structure. 
Any further discussion of these kinematic profiles would, however, be
overinterpretation.  It suffices that the basic shell appears to be
comparatively thick and extended and, as we will show in the next
section, agrees with the results of detailed photoionization models. 

The third STIS spectrum provides the strongest evidence for
homogeneity of the ejecta.  The Ne IV] 1602\AA\ and He II 1640\AA\
lines show identical profiles, with matches for each knot.  This
contrasts with our GHRS results for V1974 Cyg \citet{Sho97} where high
spatial resolution small aperture spectra show marked contrasts
between these two lines and also with C IV 1550\AA.  Closer agreement
was found for that nova between line profiles earlier in the outburst
at a time similar to those reported here, suggesting that we were not
yet completely viewing the ejecta. 

\section{Photoionization Model Analysis} 

The \cldy\ 94.00 photoionization code \citep[and references
therein]{Fer98} was used to model the observed emission line fluxes
for the August 29 observation given in Table 2.  We concentrate on
this set of spectra since by this stage the ejecta were sufficiently
optically thin, as indicated by the near identity of the line profiles
on all species.  \cldy\ simultaneously solves the equations of thermal
and statistical equilibrium for a model emission nebula.  Its output,
the predicted flux of $\sim$ 10$^4$ emission lines, is compared
against the observations to determine the physical conditions in the
shell.  \cldy\ has been used to model numerous novae (see Schwarz et
al. 1997, 2001; Vanlandingham et al. 1996, 1997, 1999). 

The outer radius of the model shell is constructed using the observed
maximum expansion velocity and the time since outburst.  From the
early P Cygni terminal velocities, we use 5200 km s$^{-1}$ and and
assume a linear velocity flow to set the outer dimension of the shell.
 The inner radius was determined from the estimated shell thickness of
0.5.  The ejecta are assumed to be spherically symmetric.  The density
of the shell is set by a hydrogen density parameter which has a power
law density profile with an exponent of -3.  This provides a constant
mass per unit volume throughout the model shell, which is a reasonable
assumption. \cldy\ also allows a filling factor of less than one. The
filling factor sets the ratio of the filled to vacuum volumes in the
ejecta.  It acts by modifying the volume emissivity and the optical
depth scale of the ejecta.  The elemental abundances are set relative
to hydrogen and we began initially with V1974 Cyg abundance solution
(Vanlandingham et al. in prep).  A hot (few 10$^6$ K) non-LTE
planetary nebula nuclei spectral energy distribution \citep{Rauch97}
with a high luminosity ($\sim$ 10$^{38}$ erg s$^{-1}$) is used as the
input source. 

Initial attempts to reproduce the observed line flux did a reasonable
job fitting the majority of the lines but failed with the highest
ionized species.  The high density and the low luminosity of the model
produced an ionization bounded shell with a hydrogen
recombination radius slightly larger than the inner radius.  As a
result, the high ionization zones in the model shell were small and
did not produce the required amount of flux.  In order to include
these other lines we added an additional, less dense component to the
previous model. The lower density means the ionizing photons 
penetrate further into the model shell, resulting in a hotter and more
ionized shell. The second component has exactly the same parameters as
the denser component except for the hydrogen density and filling
factor which was allowed to vary independently.  Table 3 gives the
comparison between the observations and the \cldy\ predictions for the
two models. Three of the lines from Table 2 may be blends based on the
\cldy\ predicted fluxes from lines of similar wavelengths.  These
lines are noted as ``Blend'' in Table 3 and all of the \cldy\ lines
within a few Angstroms are summed and their combined flux is compared
with the observation.  The fluxes are presented relative
to the \ion{He}{2} (1640\AA) line since we lack any uncontaminated
hydrogen lines in the spectrum.  The observed lines were dereddened
with E($B-V$) = 0.2.  We determined a goodness of fit from the $\chi^2$
of the model: 

\begin{equation}
\chi^2 = \sum_{i} \frac{(M_{i} - O_{i})^2}{(\sigma_{i})^2},
\end{equation}

\noindent where $O_i$ is the observed line ratio and $\sigma_{i}$ is
the error associated with the observed line ratio ($\sim$ 25\%). The
total $\chi^2$ of the combined models is $\sim$ 17 with the largest
contribution coming from the \ion{N}{5} line which is blended with the
Ly$\alpha$.  The best fit model parameters are given in Table 4.  There
are 13 free parameters in the two \cldy\ models and thus with 16 line
ratios to model there are 3 degrees of freedom.  The metal abundances
are scaled to He so if He/H$>$1, the metallicity is automatically
elevated. The derived ejected mass of the models is on the high end of
the values typically found for novae, $M_{\rm ejecta} \sim$
5$\times$10$^{-4}$ M$_{\sun}$, assuming a spherical covering factor of
one. The covering factor is the fraction of 4$\pi$ str covered by the
model shell and it scales with the \cldy\ line luminosities.  Note, a
covering factor less than unity doesn't affect the model WD luminosity
since the covering factor only scales the line luminosities associated
with the model shell. 

The derived abundances are not as extreme for V382 Vel as those we
have determined for ``fast'' novae, especially the Galactic novae V693
CrA and V838 Her, and Nova LMC 1990\# 1. Helium, carbon, and silicon
are consistent with solar abundances (He = 1.0, C = 0.6$\pm$0.3, Si =
0.5$\pm$0.3), while nitrogen, oxygen, neon, magnesium, and aluminum
are enhanced (N = 17$\pm$4, O = 3.4$\pm$0.3, Ne = 17$\pm$3, Mg =
2.6$\pm$0.1, Al = 21$\pm$2).  In general, the nitrogen, neon, and
aluminum enhancements are lower than either V693 CrA or Nova LMC 1990
\# 1.  A more detailed comparison with other ONeMg novae, including a
re-analysis of V1974 Cyg data, will be presented in a future paper. 

An independent estimate of the distance can be obtained using the the
observed \ion{He}{2} flux and the predicted \ion{He}{2} luminosity
from the combined models. The observed flux was dereddened assuming an
E($B-V$) = 0.2 and the model luminosities were calculated with a
covering factor of unity.  The distance obtained using this method,
2.8 $C^{1/2}$ kpc where $C$ is the model covering factor, is in
agreement with our previous determination of $\sim$ 2.5 kpc.   If we
use 2.5 kpc as the true distance the covering factor of the combined
models must be 0.8 which drives the ejected mass down to
4$\times$10$^{-4}$ M$_{\sun}$.  The mass derived by \citet{DV02} is
$6.5\times 10^{-6}$M$_\odot$ from the data on 2000 Oct. 2 for an
assumed distance of 1.7 kpc.  Increasing this to 2.3 kpc increases the
estimated mass to $2\times 10^{-5}$M$_\odot$, still nearly a factor of
10 below the one we derive based on the UV spectra.  However, since
this nova suffered an extended ``iron curtain'' phase in the UV, the
minimum mass must have been significantly higher than that estimated
by Della Valle et al. (see Shore 2002). 

\section{Comparisons with Nova LMC 2000} 

Nova LMC 2000 provides a comparison with V382 Vel, even more so than
V1974 Cyg.  It too was an ONeMg nova, but being situated in the LMC
provides absolute information on distance against which the Galactic
novae can be scaled.  Having observed this nova in two nearly
identical stages to those seen in V382 Vel, we here make some brief
remarks about what can be learned from the comparative developments (a
more detailed paper on Nova LMC 2000 is in preparation). 

Nova LMC 2000 was discovered by Liller on 2000 July 12.4 UT \citep{LS00}.  
Archival images \citep{DP00} show the outburst was not detected
on June 29.38 but the nova was visible a very short time later, June
29.65.  Its maximum measured visual magnitude was 11.2.  Optical spectra 
taken within two days showed emission lines of neutral and singly ionized
species, especially the Na I D lines, and P Cygni profiles with a
maximum expansion velocity of -1900 km s$^{-1}$ \citep{DP00}. 

Our STIS observations employed the same settings as those we used for
V382 Vel and occurred on 2000 August 19.7 UT and August 20.9 UT
\citep{Shore00}. Weak iron-peak absorption may still be present,
especially from 1550 to 1565\AA.  The spectrum strongly resembled the
August 29 V382 Vel spectrum but with much stronger Ly$\alpha$ and Si
III] 1895\AA\ and C III] 1910\AA\ emission, with Si III]/C III] about
2. Strong C IV P-Cyg absorption was seen with a terminal velocity of
2000 km s$^{-1}$, similar to the FWHM velocity for the emission lines.
 The strongest emission lines were C III 1076\AA, N V 1240\AA, N III/O
III 1267\AA, O I 1304\AA, C II 1334\AA, Si IV/O IV] 1400\AA, C IV
1550\AA, He II 1640\AA, O III] 1667\AA, N III] 1750\AA, Si II 1816\AA,
Al III 1860\AA, Si III] 1895\AA, C III] 1910\AA, N II] 2145\AA, C II
2321\AA, Al II 2672\AA, and Mg II 2800\AA.  Unlike V382 Vel at this
stage, there were no N IV] 1486\AA\ or 1718\AA\ lines.  The Ly$\alpha$
line remained strong and asymmetric.  A comparison of the V382 Vel and
Nova LMC 2000 Ly$\alpha$ profiles is shown in Fig. 15.  Notice that
the shift of the nova relative to the foreground Galactic contribution
is responsible for the stronger blueshifted emission for the LMC nova
and shows that the probable reason for the lack of recognition of
Ly$\alpha$ P Cyg profiles in Galactic novae is due to interstellar
absorption.  This is displayed at higher resolution in the top panel
of Fig. 15. 

The first STIS spectrum was obtained at a later stage of the outburst
than for V382 Vel.  The metallic absorption lines had already largely
disappeared, leaving strong P Cyg profiles on the UV resonance lines. 
It is important to note that this supports the conclusion drawn from
Galactic systems that ONeMg novae systematically pass through this
stage in the ultraviolet. The emission line profiles in the second
STIS spectrum, even at this relatively early stage in the outburst,
showed similar fine structure to those at the same stage in V382 Vel
and V1974 Cyg.  The integrated flux from 1150 to 3120\AA\ was $8.6
\times 10^{-11}$ erg cm$^{-2}$ s$^{-1}$, which corresponds to $5.6
\times 10^{-10}$ erg cm$^{-2}$ s$^{-1}$ corrected for a field LMC
reddening law with E(B-V) = 0.2. Adopting a distance of 52 kpc, this
corresponds to $4.4 \times 10^4$L$_\odot$ in the UV range {\it only},
since at this epoch most of the flux was being emitted in the
ultraviolet.  In view of the similarity of the spectra, these data
yield a distance for V382 Vel of 3 kpc assuming its reddening is
E(B-V) = 0.2, which is probably an upper limit for the distance of
V382 Vel. 

\section{Summary and Conclusions} 

The analysis of V382 Vel along with two other ONeMg novae, V1974 Cyg
and Nova LMC 2000, reveals a remarkable consistency in outburst
characteristics among novae of this type.  Although there is a range
of masses and abundances, the energetics and spectral development
closely follow each other.   We derive a range for E(B-V) of 0.2 to
0.3, with the most likely value being in the lower end of the range
based on a number of independent determinations, including the
comparison with the other two well observed ONeMg novae.  The maximum
expansion velocity, of $> 5000$ km s$^{-1}$, exceeds most novae of
this class (except V838 Her \citep{Van96} and LMC 1990 No. 1) and is
consistent with the ``fast'' classification.  The derived mass of 4 -
5 $\times$ 10$^{-4}$ M$_{\odot}$ based on our analysis is at the upper
end of  the range determined for fast ONeMg novae.  This mass estimate
is, however, dependent on the ejecta filling factor which is difficult
to precisely determine. Abundance enhancements relative to solar
values are found for N, Ne, Mg, and Al, while He, C, and Si are
approximately solar abundance.  In general, the enhancements are lower
than previously determined values for most ONeMg novae. Finally,
although the system geometry is unknown, profile modeling suggests it
is consistent with an inclined ring, with an angle of about $25^o$,
that should be resolvable within a few years if the distance is about
2 to 3 kpc. 

\acknowledgements 

We warmly thank Howard Lanning, Tom Ake, B-G. Anderson, and George
Sonneborn for their generous help with the STIS and FUSE observations.
We thank Massimo Della Valle for communicating electronic versions of
his ESO spectra in advance of publication and Karen Vanlandingham for
insightful discussions.  We also thank the referee for helpful
remarks.  The V382 Vel STIS spectra were obtained through an award of
Director's Discretionary Time and we thank Steve Beckwith for his
support of these observations. Support for proposal GO 8540 and GO
8671 was provided by NASA through grants from Space Telescope Science
Institute, which is operated by Associated Universities for Research
in Astronomy, Inc., under NASA contract 5-26555.  This work was
supported by STScI under programs GO 8540 and GO 8671 and by NASA
under the FUSE guest investigator program as project A093.  SNS wishes
to thank Profs. M. Salvati and F. Pacini for their kind hospitality
during extended visits to Arcetri.   SS, RDG, and CEW, and PHH
acknowledge support of NSF and NASA grants to ASU, University of
Minnesota, and University of Georgia, respectively.

\clearpage
\begin{figure}
\plotone{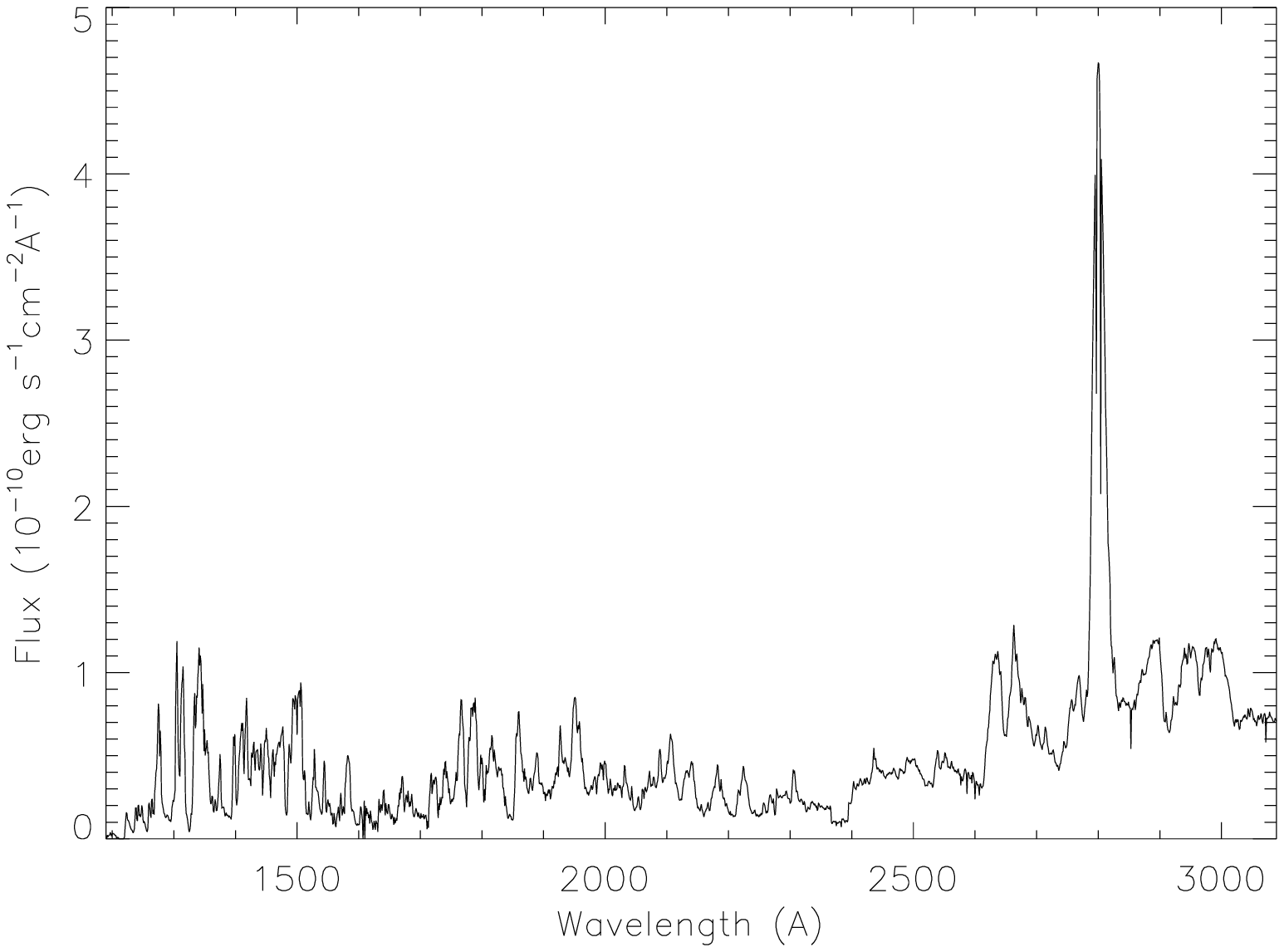}
\caption{V382 Vel, 1999 May 31 (O5JV01) observation; absolute flux 
(erg s$^{-1}$cm$^{-2}\AA^{-1}$) with 1\AA\ binning, uncorrected for 
reddening.}
\end{figure}

\clearpage
\begin{figure}
\plotone{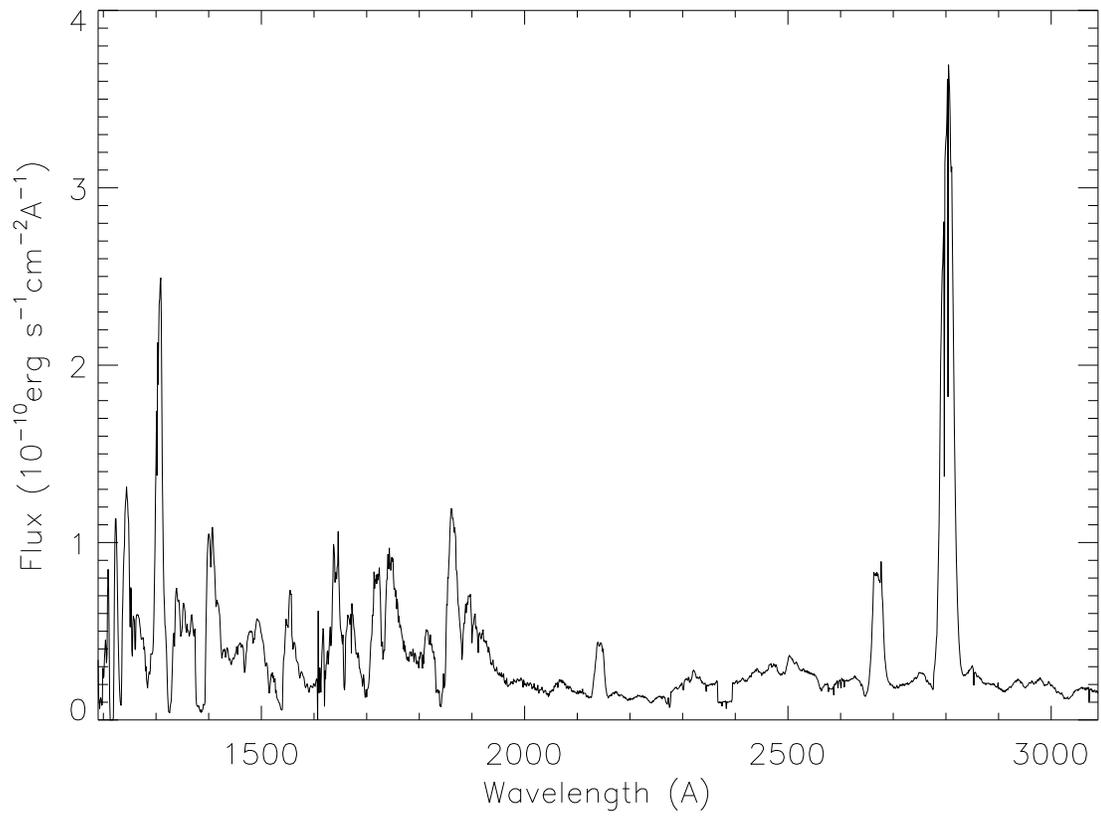}
\caption{V382 Vel, 1999 Jun 21 (O5JV02) observation; same as Fig. 1}
\end{figure}

\clearpage
\begin{figure}
\plotone{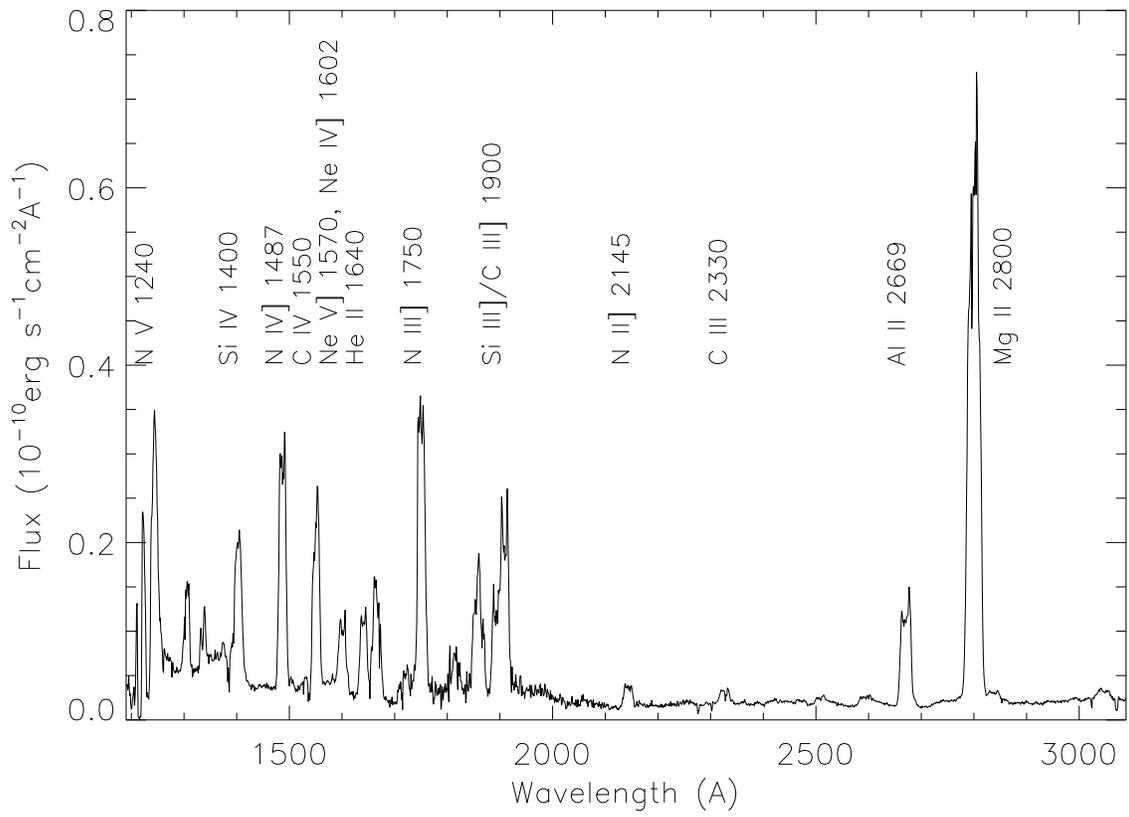}
\caption{V382 Vel, 1999 August 29 (O5JV03) observation; same as Fig. 1}
\end{figure}

\clearpage
\begin{figure}
\plotone{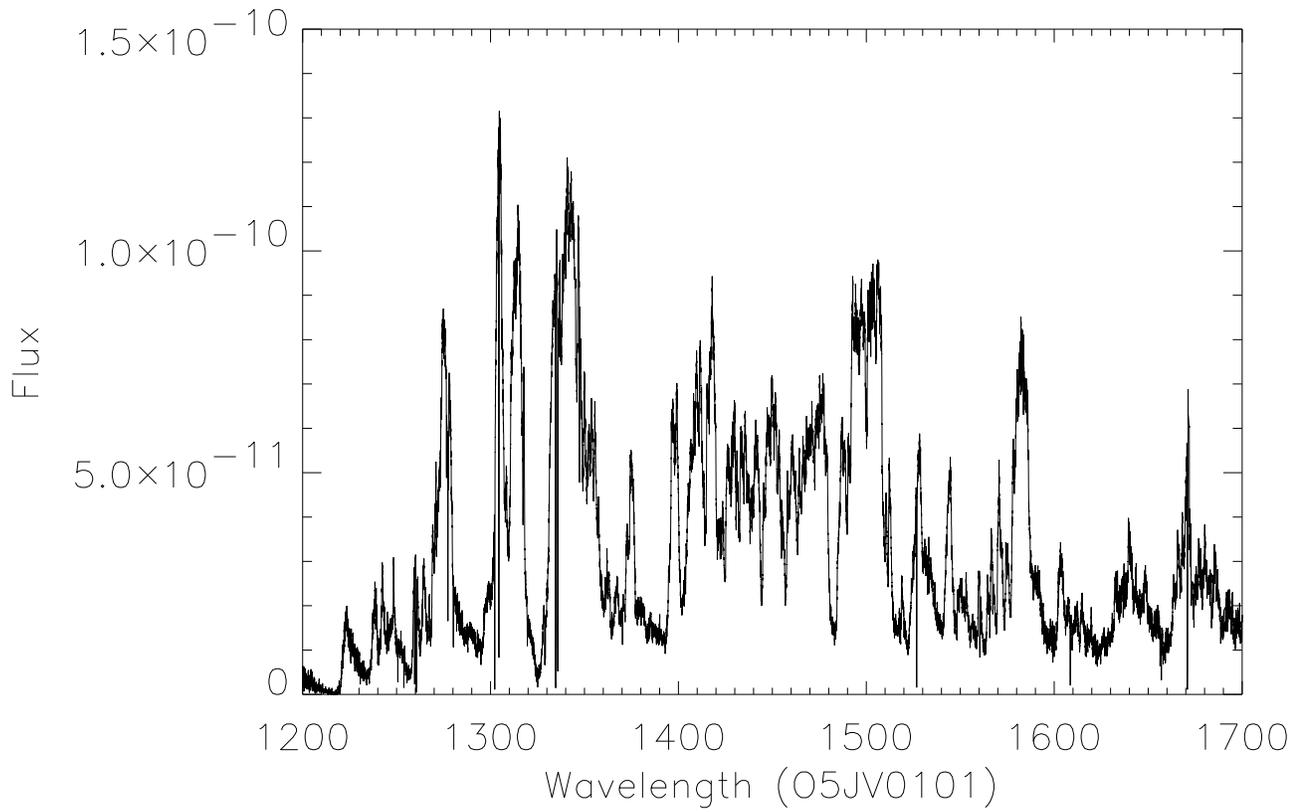}
\caption{1200 - 1700\AA\ region for 1999 May 31 (O5JV01); same as Fig. 1}
\end{figure}

\clearpage
\begin{figure}
\plotone{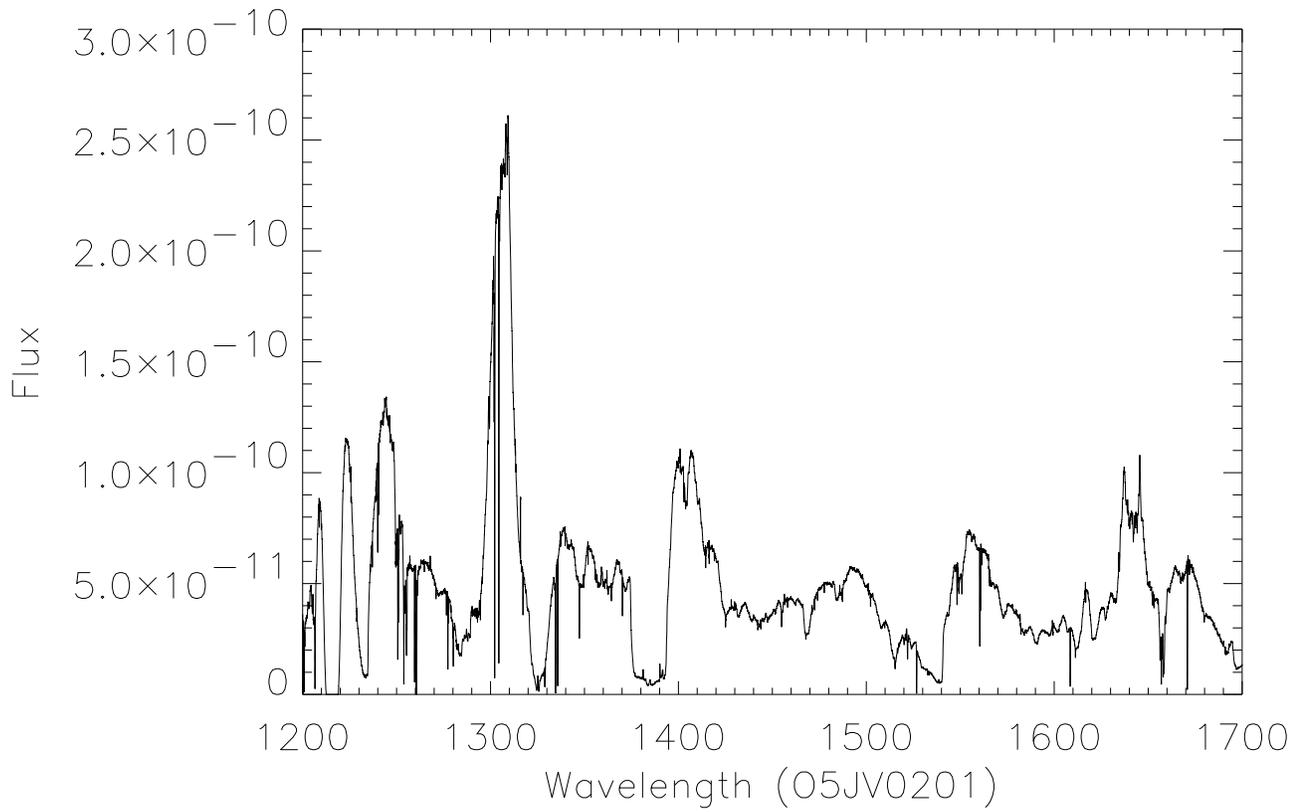}
\caption{1200 - 1700\AA\ region for 1999 Jun 21 (O5JV02); same as Fig. 1}
\end{figure}

\clearpage
\begin{figure}
\plotone{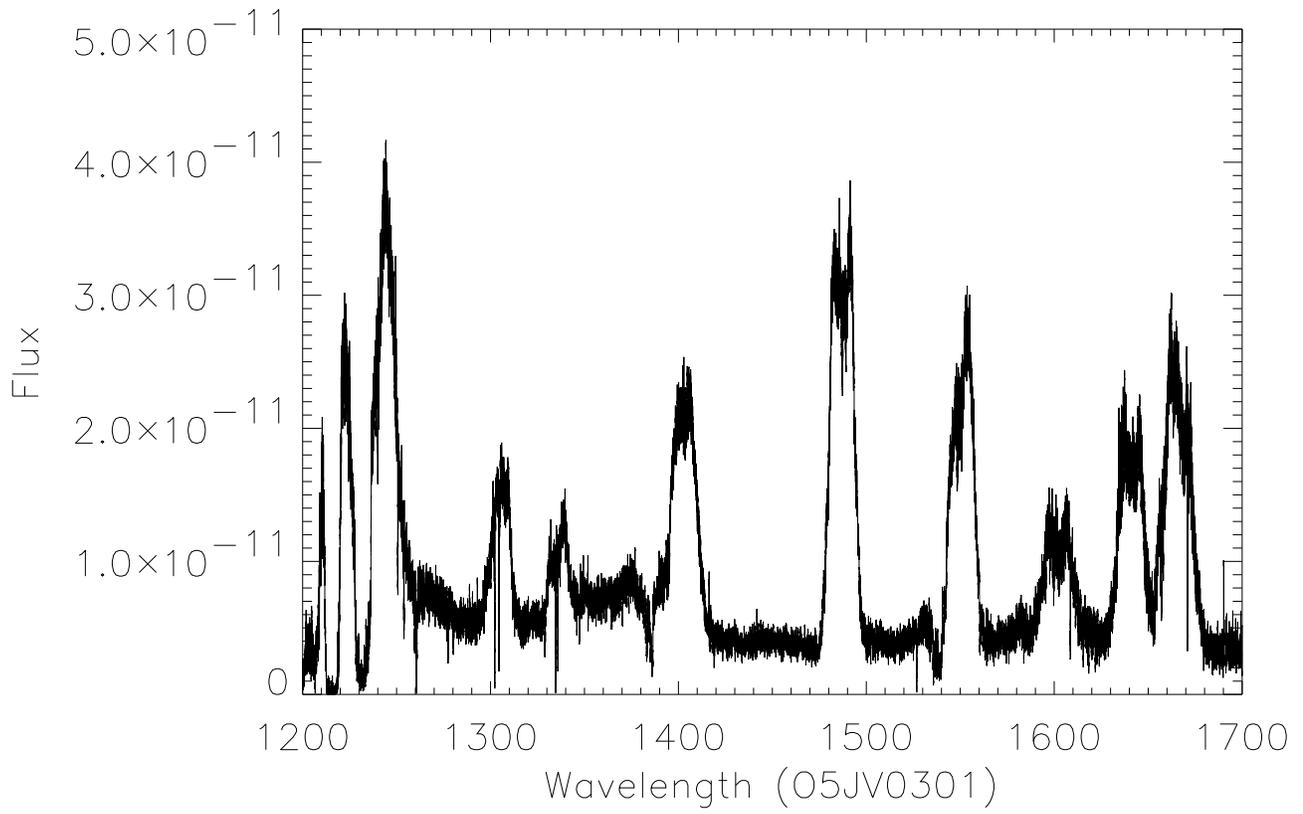}
\caption{1200 - 1700\AA\ region for August 29 (O5JV03); same as Fig. 1}
\end{figure}

\clearpage
\begin{figure}
\plotone{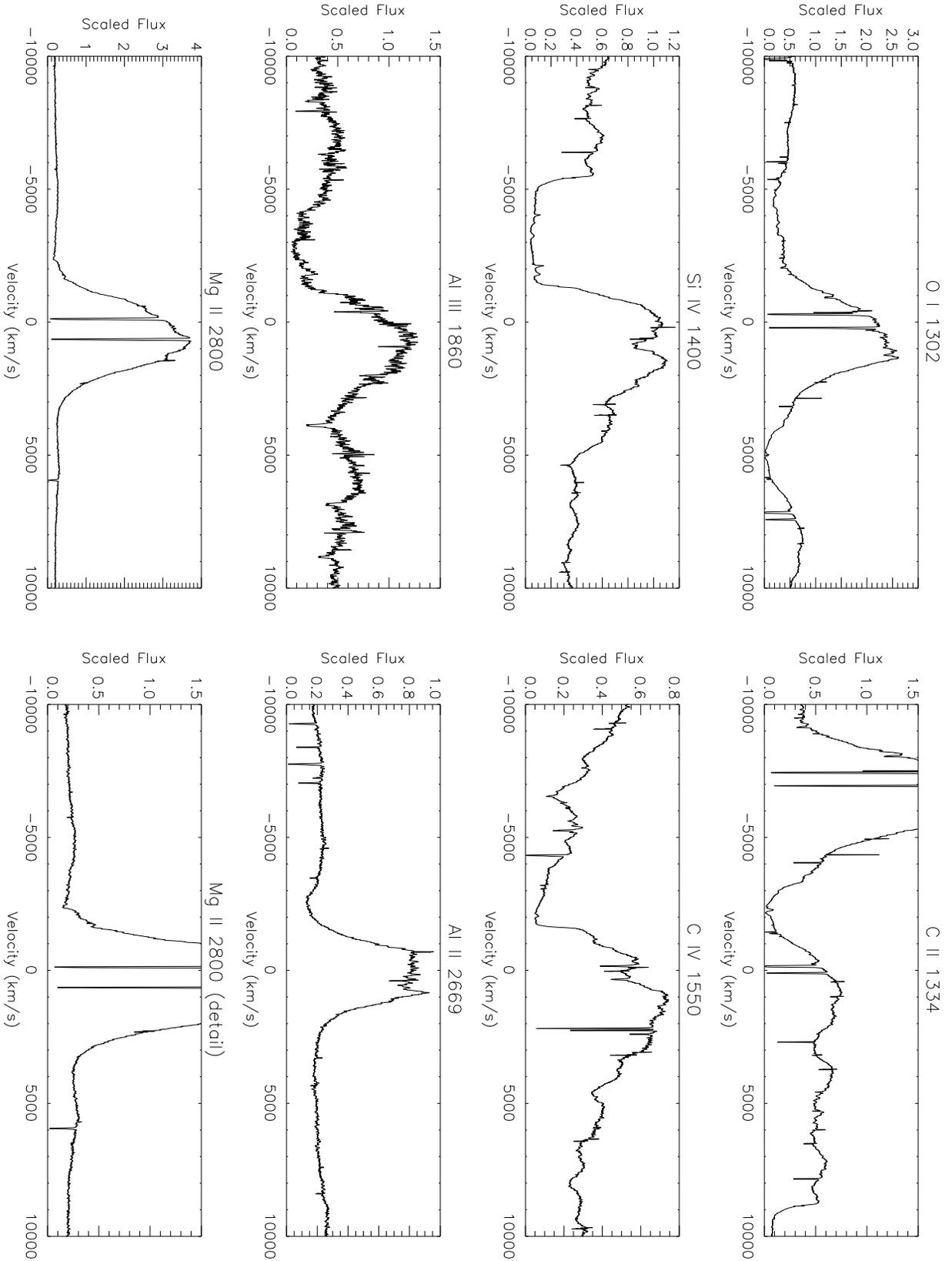}
\caption{Resonance line profiles for the Jun 21 STIS spectrum
of V382 Vel.}
\end{figure}

\clearpage
\begin{figure}
\plotone{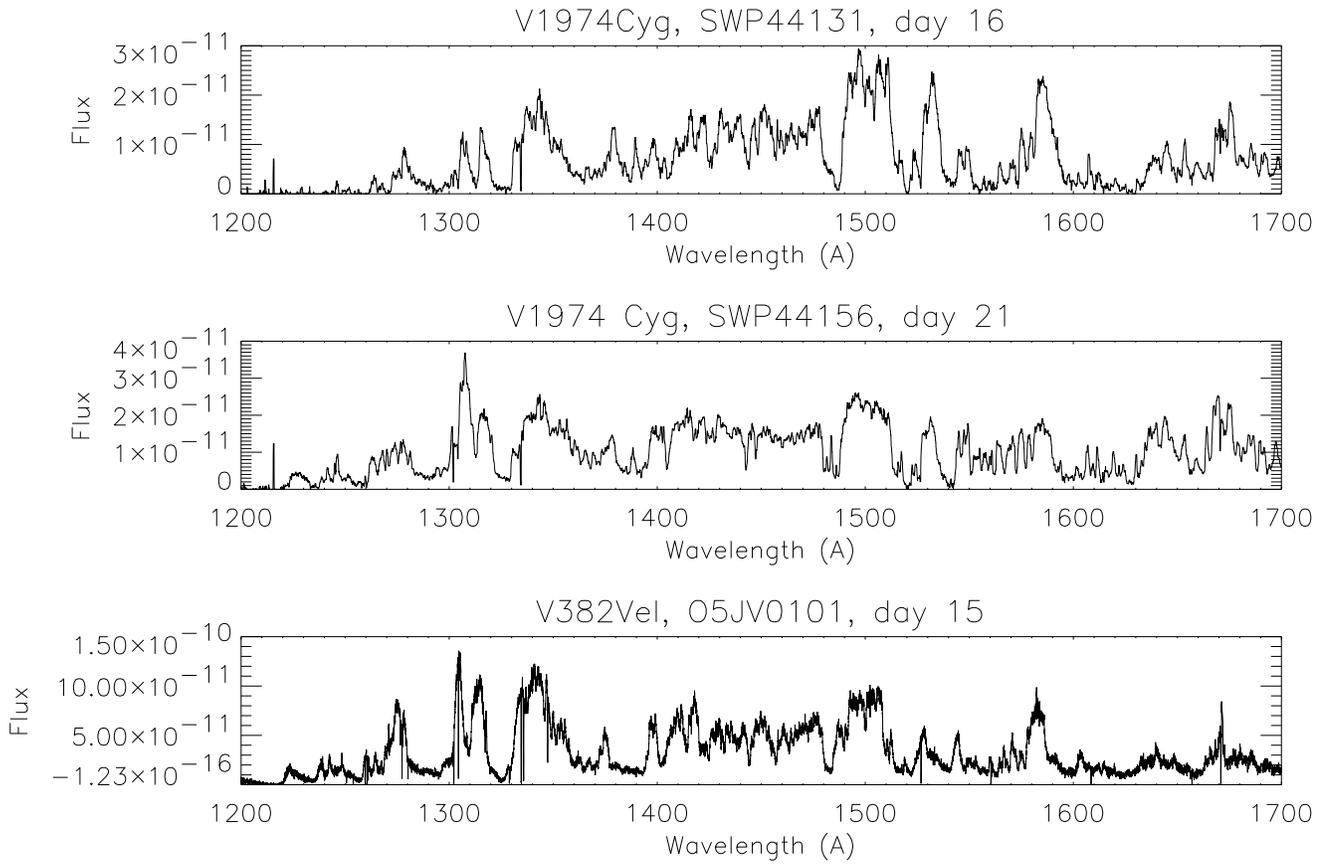}
\caption{Comparson of two early high resolution spectra of V1974 Cyg with 
the 31 May observation of V382 Vel showing likely spectral development
before the first STIS spectrum.  No reddening corrections have been 
applied.}
\end{figure}

\clearpage
\begin{figure}
\plotone{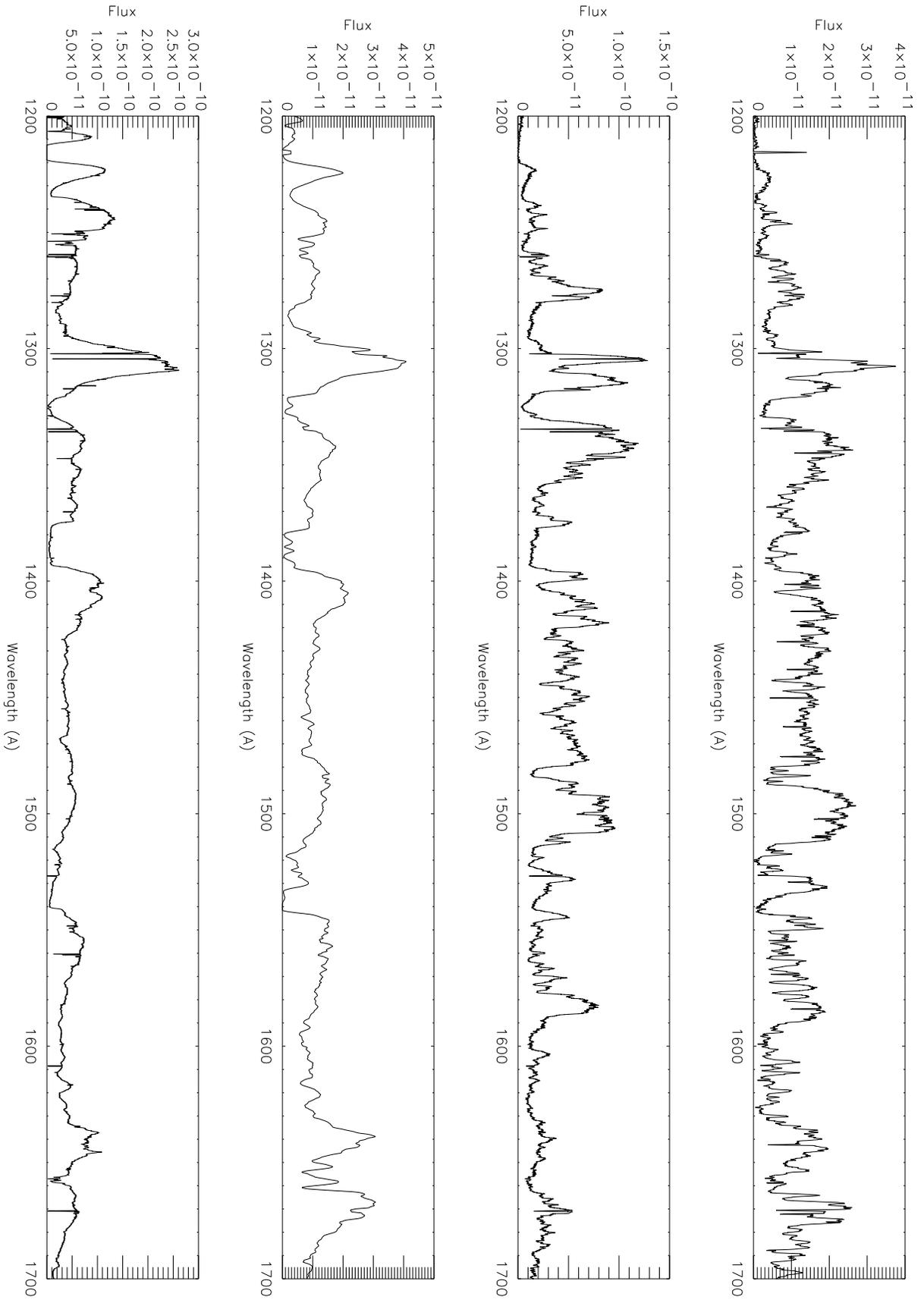}
\caption{V1974 (top), SWP 44156; V382 Vel (second), O5JV01; V1974 Cyg
(third), SWP 44378; V382 Vel (bottom), O5JV02.  No reddening 
corrections have been applied.}
\end{figure}

\clearpage
\begin{figure}
\figurenum{10A}
\plotone{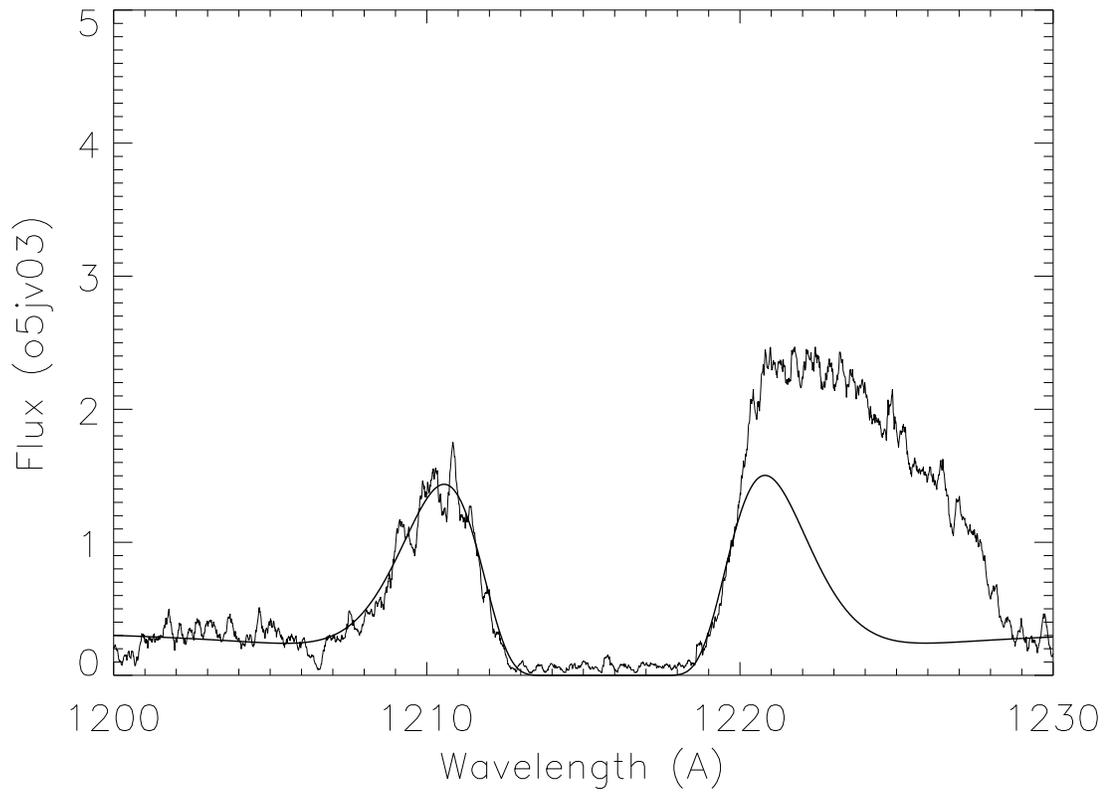}
\caption{A neutral hydrogen column density of $N_H \approx$
1.2$\times$ 10$^{21}$cm$^{-2}$ model fit to the interstellar
Ly$\alpha$ for the 1999 August 29 spectrum.} 
\end{figure}

\clearpage
\begin{figure}
\figurenum{10B}
\plotone{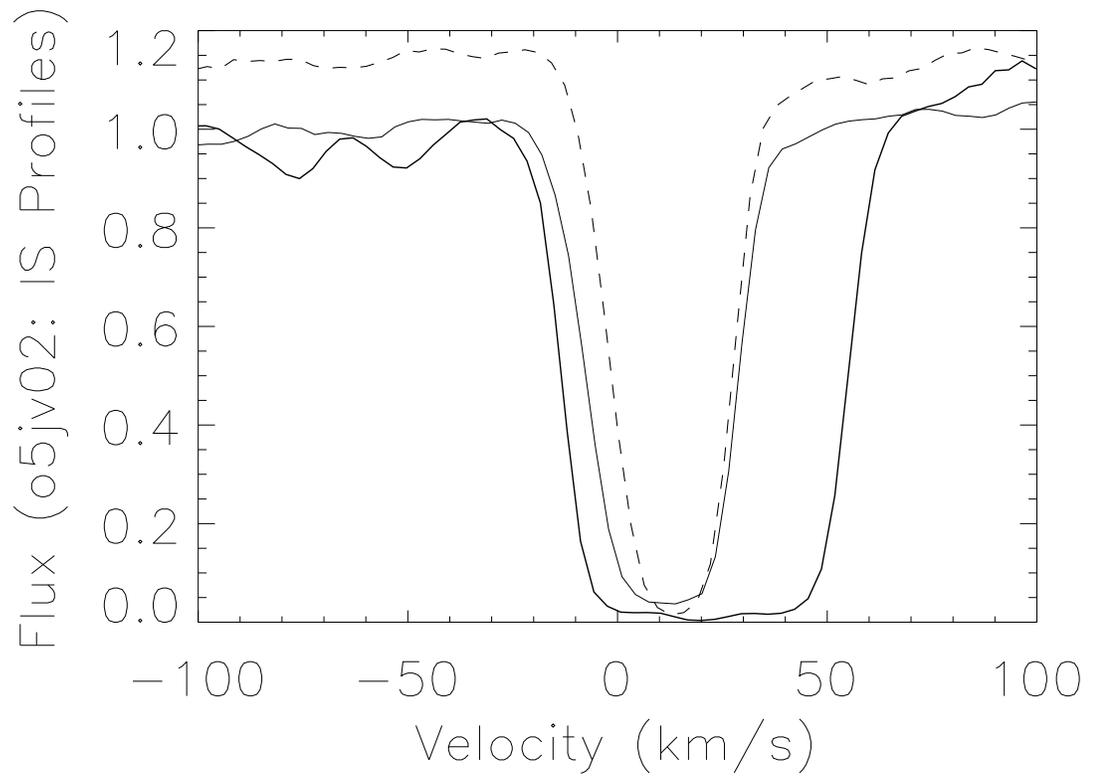}
\caption{Sample interstellar lines. The solid line shows the two
components of the C II 1334, 1335\AA\ doublet, the dashed line is Si
II 1260\AA.} 
\end{figure}

\clearpage
\begin{figure}
\figurenum{11}
\plotone{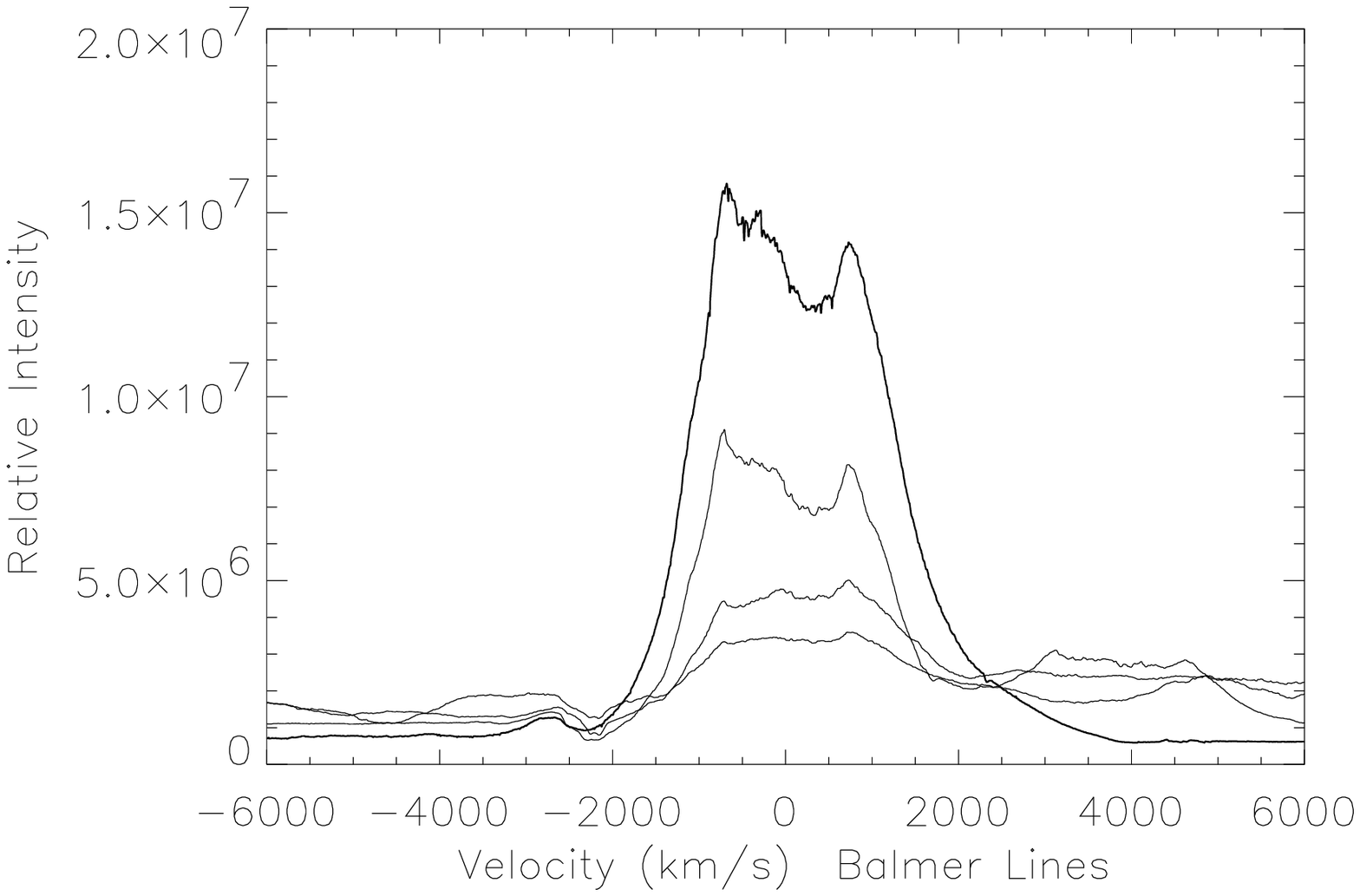}
\caption{ESO spectral line profiles: H$\alpha$ through H$\delta$ for
V382 Vel (courtesy M. Della Valle)} 
\end{figure}

\clearpage
\begin{figure}
\figurenum{12}
\plotone{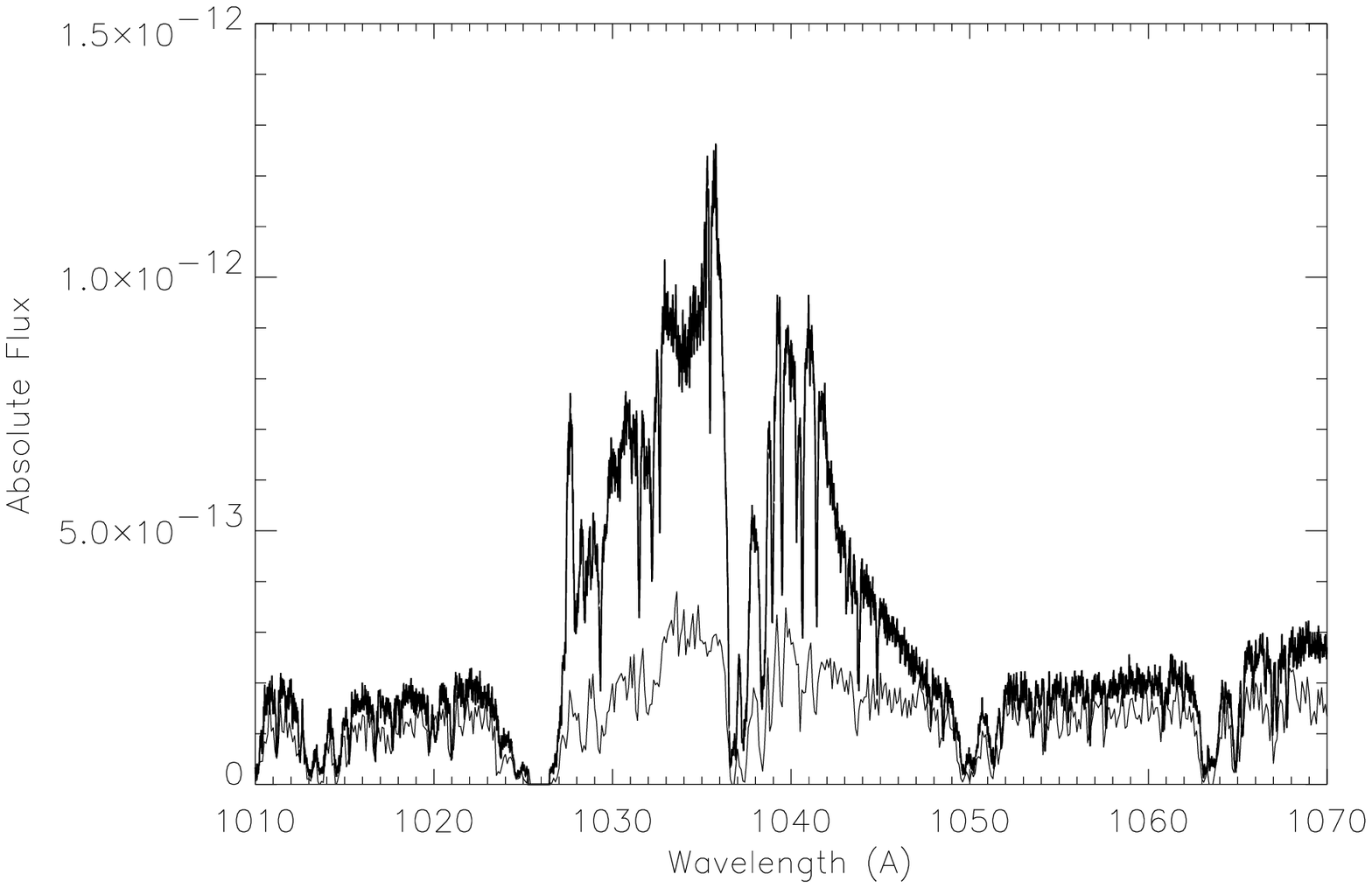}
\caption{Development of O VI 1036\AA\ doublet for V382 Vel in the two
FUSE observations (see table 1, 2000 February 6 (thick line), 2000
April 12 (thin line).  No reddening corrections have been applied, 
fluxes are in erg s$^{-1}$cm$^{-2}$\AA$^{-1}$.} 
\end{figure}

\clearpage
\begin{figure}
\figurenum{13}
\plotone{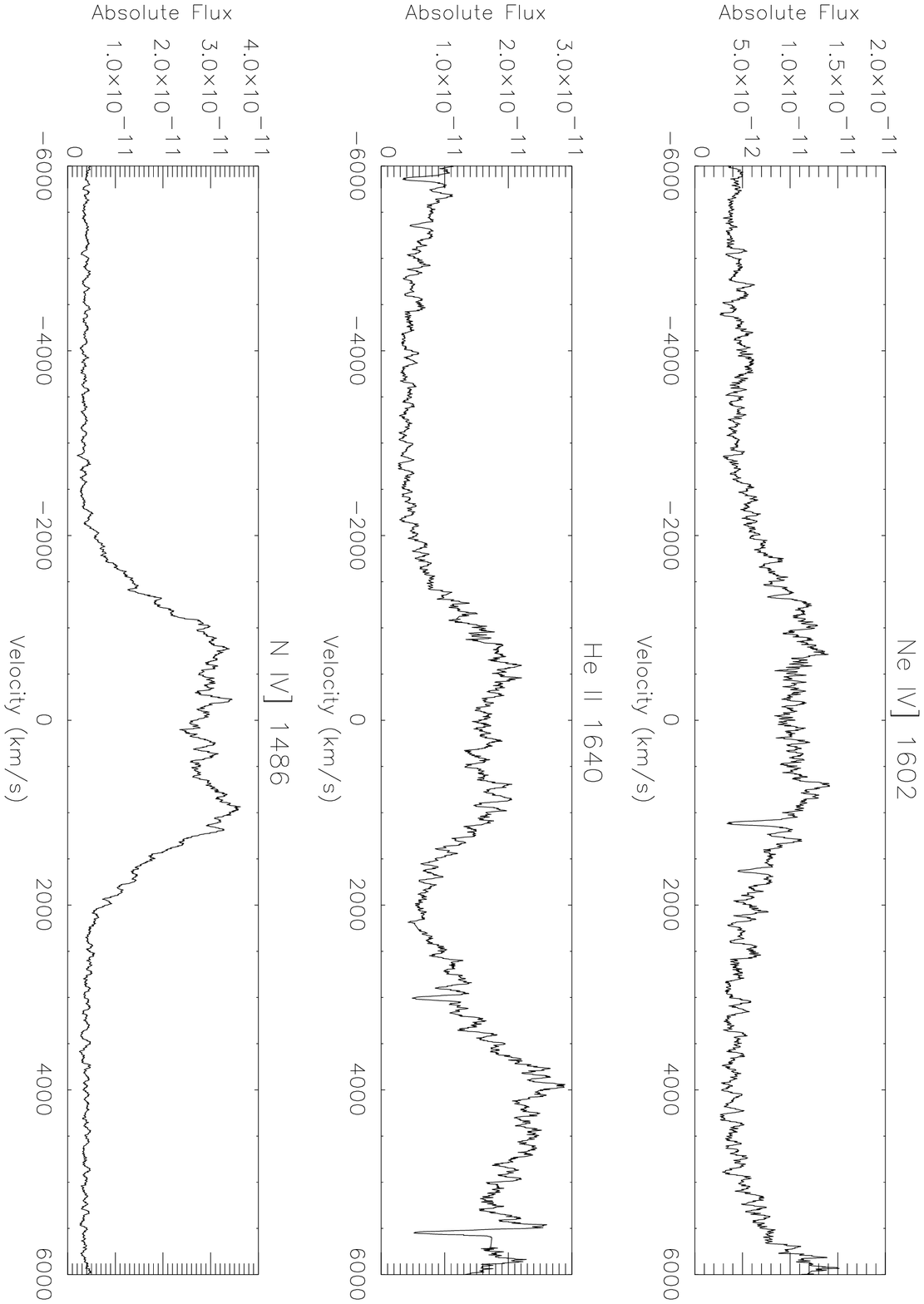}
\caption{Comparison of three emission line profiles for similar
ionization transitions of V382 Vel for 1999 August 29.} 
\end{figure}

\clearpage
\begin{figure}
\figurenum{14}
\plotone{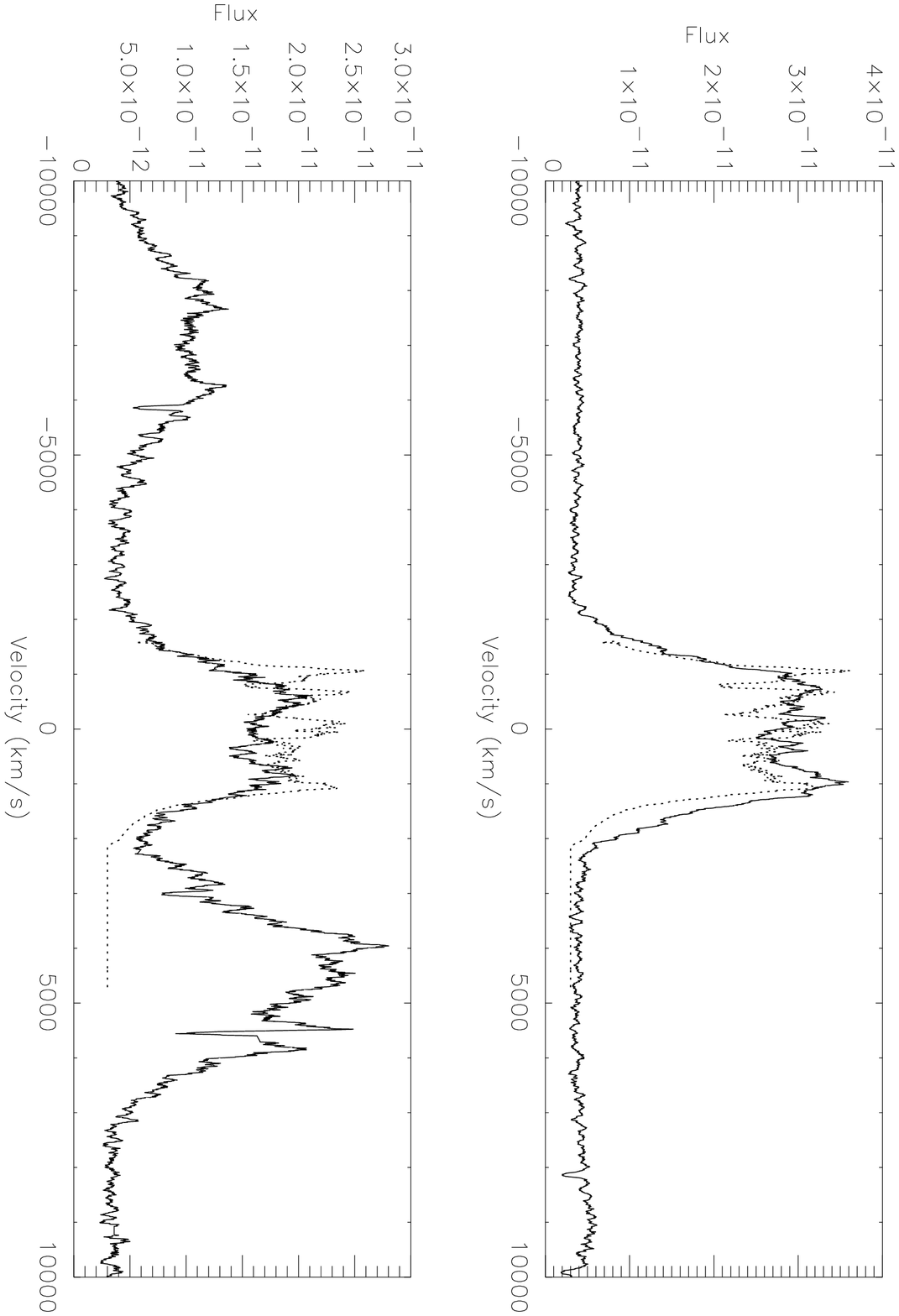}
\caption{Sample ring calculation for $\Delta R/R = 0.5$, for
$v_{\rm max} = 5200$ km s$^{-1}$ for a linear velocity law using a
quadratic density dependence for the recombination line emissivity.  
The top panel shows the comparison with the 1999 August profile of N IV] 
1486\AA, the bottom panel shows the comparison with He II 1640\AA.  
The inclination angle was 25$^{\circ}$ and the same model has been used 
in both cases.  Note the blending with O III] 1667\AA\ and the broader 
wings on Ne IV] 1602\AA\ in the bottom spectrum.}
\end{figure}

\clearpage
\begin{figure}
\figurenum{15}
\plotone{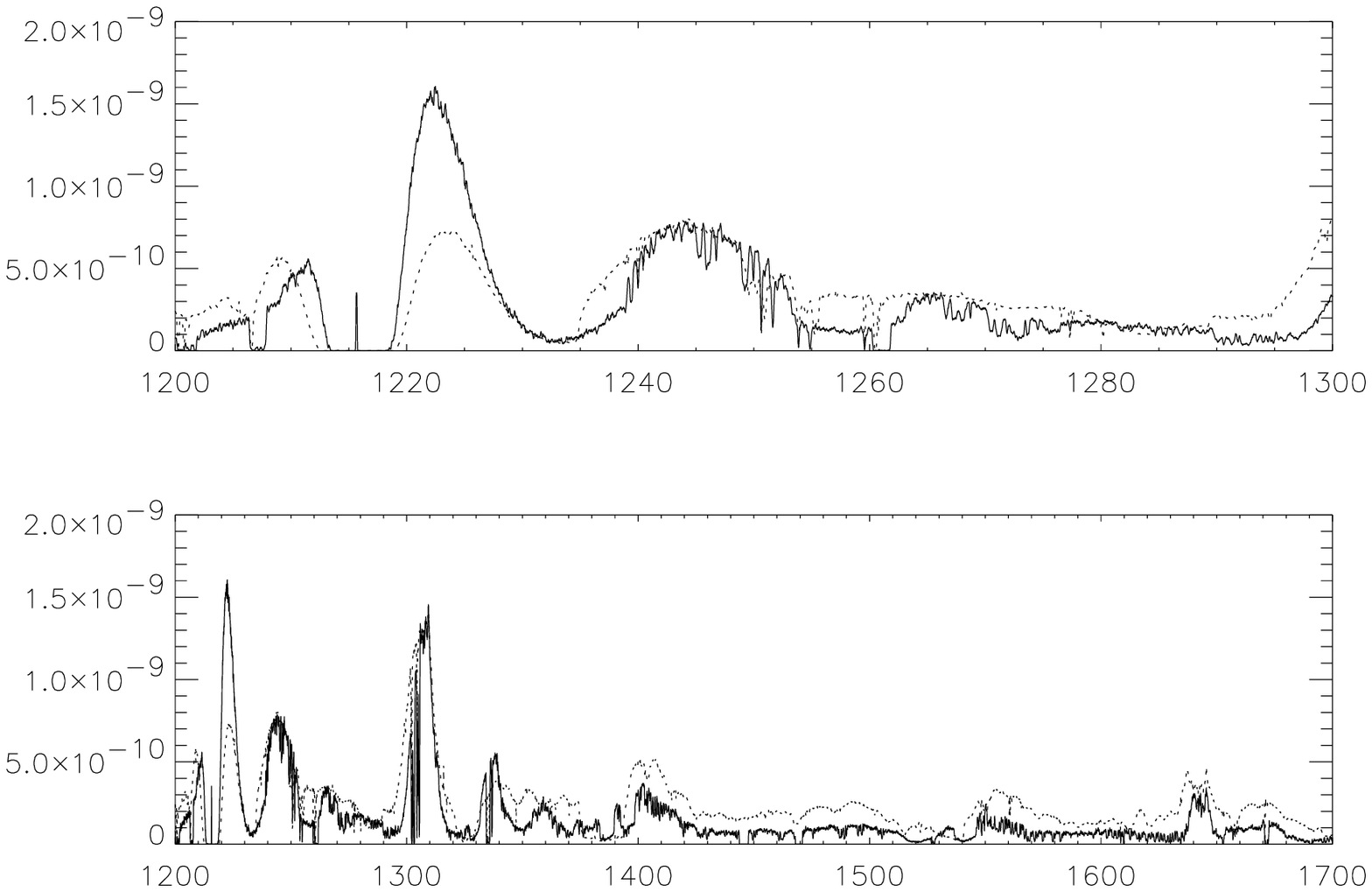}
\caption{Comparison between V382 Vel (dotted line) and Nova LMC 2000
(solid line) at similar stages in outburst, showing both spectral
similarities and the effect of different interstellar Ly$\alpha$
absorption on the appearance of the P Cyg component.  The emission at 
Ly$\alpha$ line center is geocoronal} 
\end{figure}

\clearpage

\begin{deluxetable}{ccccccc}
\tablenum{1}
\tablewidth{0pt}
\tablecaption{Journal of Observations}
\tablehead{
\colhead{Instrument} & \colhead{ID} & \colhead{Date (UT)} & \colhead{HJD} &  
\colhead{$\Delta$t\tablenotemark{a}} &
\colhead{Exp (sec)} & \colhead{$\lambda_c$ (\AA)} }
\startdata
HST/STIS & O5JV0101 & 1999 May 31.25 & 2451330 & 8 & 1680 & 1425\\
 & O5JV0102 & 1999 May 31.29 &  & & 1416 & 1978 \\
 & O5JV0103 & 1999 May 31.33 &  & & 1100 & 2707 \\
 & O5JV0201 & 1999 Jun. 21.04 & 2451351 & 29 & 1400 & 1425 \\
 & O5JV0203 & 1999 Jun. 21.08 &  & & 1430 & 1978 \\
 & O5JV0204 & 1999 Jun. 21.14 &  & & 1100 & 2707 \\
 & O5JV0301 & 1999 Aug. 29.04 & 2451420 & 98 & 1757 & 1425 \\
 & O5JV0302 & 1999 Aug. 29.11 &  & & 2875 & 1425 \\
 & O5JV0303 & 1999 Aug. 29.18 &  & & 2842 & 1978 \\
 & O5JV0304 & 1999 Aug. 29.25 & & & 2842 & 2707 \\
FUSE & A0930201 & 2000 Feb. 6 & 2451581 & 259 & 92.5$\times$10$^3$ & 1060 \\
     & A0930202 & 2000 May 4 &  2451593 & 325 & 11.6$\times$10$^3$ & 1060\\
     & A0930203 & 2000 Jul. 3 & 2451728 & 398 & 25.0$\times$10$^3$\tablenotemark{b} & 1060 \\
     & A0930204 & 2000 Jul. 3 & 2451728 & & - & 1060 \\
\enddata
\tablenotetext{a}{Time since visual maximum (1999 May 23.3 UT).}
\tablenotetext{b}{Combined exposure time.}
\end{deluxetable}

\begin{deluxetable}{cccc}
\tablenum{2}
\tablewidth{0pt}
\tablecaption{Emission Line Strengths for STIS Spectrum O5JV03}
\tablehead{
\colhead{Species} & \colhead{Wavelength} & \colhead{Flux} & \colhead{Notes}\\
\colhead{} & \colhead{\AA} & \colhead{$10^{-10}$ erg s$^{-1}$cm$^{-2}$} & 
\colhead{} }
\startdata
N V & 1240 & 3.00 & a \\
O I & 1302 & 1.01 &  \\
C II & 1335 & 0.50 & \\
O V & 1375 & 0.13 & \\
Si IV & 1400 & 2.41 & \\
N IV$]$ & 1486 & 3.76 & \\
C IV & 1550 & 2.34 & \\
$[$Ne IV$]$ & 1602 & 1.17 & b \\
He II & 1640 & 2.07 & c \\
O III$]$ & 1667 & 3.06 & \\
N III$]$ & 1750 & 4.99 & \\
Si II & 1816 & 0.56 & \\
Al III & 1860 & 2.41 & \\
Si III$]$+C III$]$ & 1900 & 4.37 & d \\
N II & 2147 & 0.42 &  \\
C III & 2321 & 0.31 & \\
Al II & 2670 & 2.41 & \\
Mg II & 2800 & 14.3 & \\
O III & 3045 & 0.28 &  \\
\enddata
\tablenotetext{a}{No extinction corrections have been applied 
to any quoted fluxes.}
\tablenotetext{b}{P Cyg profile, blended with Ly $\alpha$}
\tablenotetext{c}{ Upper limit for Ne V] 1575\AA\ 
is 2$\times 10^{-11}$ erg s$^{-1}$cm$^{-2}$; 
(c) blend on red wing with O III]; (d) F(Si III])/F(C III]) = 0.3. } 
\end{deluxetable}

\begin{deluxetable}{lcccccc}
\tablenum{3}
\tablewidth{0pt}
\tablecaption{{\it Cloudy} Emission Line Predictions}
\tablehead{
\colhead{Ion} & \colhead{Wavelength} & 
\colhead{1st Component\tablenotemark{a}} & 
\colhead{2nd Component\tablenotemark{a}} &
\colhead{Total\tablenotemark{a}} &
\colhead{Observed\tablenotemark{b}} &
\colhead{$\chi^2$} \\
\colhead{} & \colhead{(\AA)} & \colhead{(Dense)} & \colhead{(Hot)} &
\colhead{} & \colhead{} & \colhead{}
}
\startdata
\ion{N}{5}     & 1240 & 0.088  &  0.674  &  0.762  &  2.00  &  6.133\\
Blend          & 1400 & \nodata & \nodata & \nodata & 1.26 &   3.975 \\
\ion{Si}{4}    & 1397 & 0.091  &  0.019  &  0.110   & \nodata& \nodata \\
\ion{O}{4}     & 1402 & 0.263  &  0.260  &  0.522   & \nodata& \nodata \\
\ion{N}{4}]    & 1486 & 1.056  &  0.854  &  1.910   & 1.87 &   0.438 \\
\ion{C}{4}     & 1549 & 0.763  &  0.436  &  1.199   & 1.15 &   1.023 \\
\ion{Ne}{5}]   & 1575 & 0.003  &  0.038  &  0.042   & $<$0.10 & \nodata \\
\[\ion{Ne}{4}] & 1602 & 0.146  &  0.424  &  0.570   & 0.57 &   0.060 \\
\ion{He}{2}    & 1640 & 0.507\tablenotemark{c}   & 
0.493\tablenotemark{d}   & 1.000   & 1.00 &   0.000 \\
\ion{O}{3}]    & 1665 & 1.411  &  0.097  &  1.507  &  1.46  &  0.017\\
\ion{N}{3}]    & 1750 & 2.416  &  0.079  &  2.495  &  2.31  &  0.103\\
Blend          & 1810 & \nodata & \nodata & \nodata & 0.26  & 1.226\\
\ion{Si}{2}    & 1808 & 0.048  &  0.000  &  0.048  & \nodata& \nodata \\
\ion{Ne}{3}    & 1815 & 0.127  &  0.013  &  0.140  & \nodata& \nodata \\
\ion{Al}{3}    & 1860 & 0.873  &  0.026  &  0.899  &  1.13  &  0.670\\
\ion{Si}{3}]   & 1888 & 0.379  &  0.005  &  0.384  &  0.49  &  0.754\\
\ion{C}{3}]    & 1909 & 1.213  &  0.030  &  1.243  &  1.63  &  0.903\\
\ion{N}{2}     & 2140 & 0.294  &  0.000  &  0.295  &  0.26  &  0.282\\
Blend          & 2324 & \nodata & \nodata & \nodata & 0.17 &   2.749 \\
\ion{O}{3}     & 2321 & 0.070  &  0.031  &  0.100   & \nodata& \nodata \\
\ion{C}{2}     & 2326 & 0.140  &  0.000  &  0.140   & \nodata& \nodata \\
\ion{Al}{2}    & 2665 & 1.036  &  0.001  &  1.037   & 0.88  &  0.511\\
\ion{Mg}{2}    & 2798 & 4.675  &  0.030  &  4.705   & 4.83  &  0.011\\
\enddata
\tablenotetext{a}{Flux relative to the sum of the two \ion{He}{2} fluxes}
\tablenotetext{b}{Dereddened with E($B-V$) = 0.2 and relative to \ion{He}{2}}
\tablenotetext{c}{\ion{He}{2} Luminosity = 4.3$\times$10$^{35}$ erg s$^{-1}$}
\tablenotetext{d}{\ion{He}{2} Luminosity = 4.2$\times$10$^{35}$ erg s$^{-1}$}
\end{deluxetable}

\newpage

\begin{deluxetable}{lc}
\tablenum{4}
\tablewidth{0pt}   
\tablecaption{{\it Cloudy} Model Parameters}
\tablehead{
\colhead{Parameter} & \colhead{Day 110}
}
\startdata
T$_{eff}$ & 1.5$\times$10$^5$ K  \\ 
Source luminosity & 5$\times$10$^{37}$ erg s$^{-1}$ \\
Hydrogen density & 1.26$\times$10$^8$, 1.26$\times$10$^7$ cm$^{-3}$ \\
Inner radius\tablenotemark{a} & 2$\times$10$^{15}$ cm \\
Outer radius\tablenotemark{a} & 5$\times$10$^{15}$ cm \\
filling factor & 0.05, 0.1 \\
He/He$_{\sun}$\tablenotemark{b} & 1.0 (1) \\
C/C$_{\sun}$\tablenotemark{b} & 0.6 (3) \\
N/N$_{\sun}$\tablenotemark{b} & 17 (4) \\ 
O/O$_{\sun}$\tablenotemark{b} & 3.4 (3) \\ 
Ne/Ne$_{\sun}$\tablenotemark{b} & 17 (3) \\
Mg/Mg$_{\sun}$\tablenotemark{b} & 2.6 (1) \\
Al/Al$_{\sun}$\tablenotemark{b} & 21 (2) \\
Si/Si$_{\sun}$\tablenotemark{b} & 0.5 (3) \\
\enddata
\tablecomments{The first number provided in the density and filling factor
rows is from the dense (1st) component while the second number is from
the hot (2nd) component. The number in the parentheses in the abundance
rows indicates the number of {\it Cloudy} lines used in the analysis.}
\tablenotetext{a}{Calculated assuming a maximum expansion velocity of 5400 
km s$^{-1}$ and a ring thickness of 0.5.}
\tablenotetext{b}{Where Log(Solar number abundances relative to hydrogen)
He:-1.0 C: -3.45 N:-4.03 O: -3.13 Ne: -3.93 Mg: -4.42 Al: -5.53
Si: -4.45 \citep{GN93}.}
\end{deluxetable}

\newpage

\begin{deluxetable}{ccc}
\tablenum{5}
\tablewidth{0pt}   
\tablecaption{Interstellar Lines: Equivalent Widths}
\tablehead{
\colhead{Ion} & \colhead{$\lambda_{\rm lab}$ (\AA)} & \colhead{EW (m\AA)}
}
\startdata
N I & 1199 & 234 \\
    & 1200 & 172 \\
    & 1201 & 150 \\
S II & 1259 & 138 \\
Si II & 1260 & 468 \\
P II & 1301 & 38 \\
O I & 1302 & 221 \\
    & 1304 & 175 \\
    & 1306 & $<$3 \\
C II & 1334 & 307 \\
     & 1335 & 163 \\
Si IV & 1400 & 9 \\
Si II & 1526 & 236 \\
P II & 1533 & 25 \\
C IV & 1548 & 65 \\
     & 1550 & 37 \\
Al II & 1671 & 248 \\
Al III & 1854 & 145 \\
       & 1863 & 42 \\
Mg II & 2796 & 575 \\
      & 2804 & 510 \\
\enddata
\end{deluxetable}


\clearpage

\begin{thebibliography}{}

\bibitem[Austin et al.(1996)]{Austin96} Austin, S.J., Wagner, R. M.,
Starrfield, S., Shore, S. N., Sonneborn, G., \& Bertram, R.\ 1996,
\aj, 111, 869 

\bibitem[Brand \& Blitz(1993)]{Brand93} Brand, J.~\& Blitz, L.\ 1993,
\aap, 275, 67 

\bibitem[Burton(1985)]{Burton85} 
Burton, W. B. 1985, \aaps, 62, 365

\bibitem[Dame et al.(1987)]{Dame87} Dame, T. M. et al.\ 1987, \apj,
322, 706 

\bibitem[Dame(1999)]{Dame99} Dame, T. M.\ 1999, The Physics and
Chemistry of the Interstellar Medium, Proceedings of the 3rd
Cologne-Zermatt Symposium, held in Zermatt, September 22-25, 1998,
Eds.: V. Ossenkopf, J. Stutzki, and G. Winnewisser, GCA-Verlag
Herdecke, ISBN 3-928973-95-9, 100 

\bibitem[Della Valle, Pasquini, \& Williams(1999)]{DV99} 
Della Valle, M., Pasquini, L., \& Williams, R.\ 1999, \iaucirc, 7193, 1

\bibitem[Della Valle et al.(2002)]{DV02} Della Valle, M, Pasquini, L.,
Daon, D., \& Williams, R. E. 2002, \aap, 390, 155 

\bibitem[Dfraine \& Tan(2002)]{DT02} Draine, B. T., \& Tan, J. C. 
2002, {\it astro-ph/0208302}

\bibitem[Duerbeck \& Pompei(2000)]{DP00} Duerbeck, H.~W.~\& 
Pompei, E.\ 2000, \iaucirc, 7457, 1

\bibitem[Ferland et al.(1998)]{Fer98} Ferland, G.J., Korista, K.T.,
Verner, D.A., Ferguson, J.W., Kingdon, J.B. \& Verner, E.M. 1998,
\pasp, 110, 761 

\bibitem[Grevesse \& Noel(1993)]{GN93} Grevesse, N. \& Noels, A.,
1993, in Prantzos, de. N., Vangioni-Flam, E., \& Casse, M., eds,
Origin \& Evolution of the Elements. Cambridge Univ. Press, Cambridge,
p. 15 

\bibitem[Hayward et al.(1996)]{Hay96} Hayward, T. L. et al.\ 1996,
\apj, 469, 854 

\bibitem[Hidayat, Ikbal Arifyanto, Aria Utama, \& 
Athiya(1999)]{1999IAUC.7188....2H} Hidayat, B., Ikbal Arifyanto, M., Aria 
Utama, J., \& Athiya, S.\ 1999, \iaucirc, 7188, 2

\bibitem[Lee et al.(1999)]{Lee99} Lee, S., Pearce, A., 
Gilmore, C., Pollard, K.~R., McSaveney, J.~A., Kilmartin, P.~M., \& 
Caldwell, P.\ 1999, \iaucirc, 7176, 1 

\bibitem[Liller \& Stubbings(2000)]{LS00} Liller, W.~\& 
Stubbings, R.\ 2000, \iaucirc, 7453, 1 

\bibitem[Morton(1991)]{Mor91} Morton, D. C.\ 1991, \apjs, 77, 119 

\bibitem[Mukai \& Ishida(2001)]{Mukai2001} Mukai, K. \& Ishida, M.\
2001, \apj, 551, 1024 

\bibitem[Orio et al.(2001)]{Orio2001} Orio, M. et al.\ 2001, \mnras,
326, L13 

\bibitem[Payne-Gaposhkin(1957)]{Pay57} Payne-Gaposchkin, C. 1957, The
Galactic Novae, New York, Dover 

\bibitem[Platais et al.(2000)]{P00} Platais, I., Girard, T. M.,
Kozhurina-Platais, V., van Altena, W. F., Jain, R. K., \& L{\' o}pez,
C. E.\ 2000, \pasp, 112, 224 

\bibitem[Paresce, Livio, Hack, \& Korista(1995)]{Paresce95} 
Paresce, F., Livio, M., Hack, W., \& Korista, K.\ 1995, \aap, 299, 823

\bibitem[Rauch(1997)]{Rauch97} Rauch, T. 1997, \aap, 320, 237 

\bibitem[Schwarz et al.(1997)]{Sch97} Schwarz, G.J., Starrfield, S.,
Shore, S.N., \& Hauschildt, P.H. 1997, \mnras, 290, 75 

\bibitem[Schwarz et al.(2001)]{Sch01} Schwarz, G.J., Shore, S.N.,
Starrfield, S., Hauschildt, P.H., Della Valle, M., \& Baron, E. 2001,
\mnras, 320, 103 

\bibitem[Shore et al.(1993)]{Sho93} Shore, S.N., Sonneborn, G.,
Starrfield, S., Gonzalez-Riestra, R., \& Ake, T.B. 1993, AJ, 106, 2408

\bibitem[Shore et al.(1994)]{Sho94} Shore, S.N., Sonneborn, G.,
Starrfield, S., Gonzalez-Riestra, R., \& Polidan, R.S. 1994, AJ 421,
344 

\bibitem[Shore(2002)]{Shore02} Shore, S. N. 2002, International Conference 
on Classical Nova Explosions, Eds.: Hernanz J. and Jos\`e, J. (NY: AIP 
Press), p. 175

\bibitem[Shore \& Starrfield(1998)]{Sho98} Shore, S. N. \& Starrfield,
S.\ 1998, Stellar Evolution, Stellar Explosions and Galactic Chemical
Evolution, 413 

\bibitem[Shore et al.(2000)]{Shore00} Shore, S.~N.~et al.\ 
2000, \iaucirc, 7486, 1

\bibitem[Steiner, Campos, \& Cieslinski(1999)]{SCC99} 
Steiner, J.~E., Campos, R., \& Cieslinski, D.\ 1999, \iaucirc, 7185, 2

\bibitem[Vanlandingham et al.(1996)]{Van96} Vanlandingham K.M.,
Starrfield S., Wagner R.M., Shore S.N., \& Sonneborn G. \mnras, 1996,
282, 563 

\bibitem[Vanlandingham et al.(1997)]{Van97} Vanlandingham K.M.,
Starrfield S., \& Shore S.N. 1997, \mnras, 290, 87 

\bibitem[Vanlandingham et al.(1999)]{Van99} Vanlandingham K.M.,
Starrfield S., \& Shore S.N. 1999, \mnras, 308, 577 

\bibitem[Warner(1987)]{War87} Warner, B. 1987, \mnras, 227, 23

\bibitem[Woodward, Wooden, Pina, \& Fisher(1999)]{Wood99} 
Woodward, C.~E., Wooden, D.~H., Pina, R.~K., \& Fisher, R.~S.\ 1999, 
\iaucirc, 7220, 3

\end{thebibliography}
\end{document}